\documentclass[a4paper,11pt]{article}
\pdfoutput=1 

\usepackage{jheppub} 

\usepackage[T1]{fontenc} 

\usepackage{amsmath}
\usepackage{amssymb}
\usepackage{amsthm}
\usepackage{physics}
\usepackage{bbm}
\usepackage{enumitem}
\usepackage{mathrsfs}

\DeclareFontFamily{OT1}{pzc}{}
\DeclareFontShape{OT1}{pzc}{m}{it}{<-> s * [1.10] pzcmi7t}{}
\DeclareMathAlphabet{\mathpzc}{OT1}{pzc}{m}{it}

\newtheorem{theorem}{Theorem}
\newtheorem{lemma}{Lemma}
\newtheorem*{lemma*}{Lemma}

\newtheorem{corollary}{Corollary}

\theoremstyle{definition}
\newtheorem{definition}{Definition}

\theoremstyle{remark}
\newtheorem{remark}{Remark}

\title{The holographic map as a conditional expectation}

\author{Thomas Faulkner}

\affiliation{University of Illinois at Urbana-Champaign$^\natural$ and Kavli Institute for Theoretical Physics}

\affiliation{$\left.\right.^{\natural}$Department of Physics, University of Illinois at Urbana-Champaign, IL USA }
\emailAdd{tomf@illinois.edu}
\abstract{We study the holographic map 
in AdS/CFT, as modeled by a quantum error correcting code with exact complementary recovery.
We show that the map is determined by local conditional expectations acting on the operator algebras of the boundary/physical Hilbert space.
Several existing results in the literature follow easily
from this perspective. The Black Hole area law, and more generally the Ryu-Takayanagi area operator, arises from a central sum of entropies on the relative commutant. These entropies are determined in a state independent way by the conditional expectation. The conditional expectation can also be found via a minimization procedure, similar to the minimization involved in the RT formula.
For a local net of algebras associated to connected boundary regions, we show the complementary recovery condition is equivalent to the existence of a standard net of inclusions -- an abstraction of the mathematical structure governing QFT superselection sectors given by Longo and Rehren. 
For a code consisting of algebras associated to two disjoint regions
of the boundary theory we impose an extra condition, dubbed dual-additivity, that gives rise to phase transitions between different entanglement wedges.  
Dual-additive codes naturally give rise to a new split code subspace, and an entropy bound controls which subspace and associated algebra is reconstructable. We also discuss known shortcomings of exact complementary recovery  as a model of holography.
For example, these codes are not able to accommodate holographic violations of additive for overlapping regions.   
We comment on how approximate codes can fix these issues.
}
\begin{document}

\maketitle
\flushbottom

\section{Introduction}

The connection between quantum error correcting codes and quantum gravity \cite{Almheiri:2014lwa,Dong:2016eik,Harlow:2016vwg} is the culmination of various important insights into the nature of gravity that started with black hole thermodynamics \cite{Bekenstein:1973ur,Hawking:1974sw} went through the holographic principle \cite{'tHooft:1993gx,Susskind:1994vu}, AdS/CFT \cite{Maldacena:1997re,Gubser:1998bc,Witten:1998qj,Susskind:1998dq} and in particular hinged greatly on the generalization of the Bekenstein-Hawking entropy in the Ryu-Takayangi (RT) formula \cite{Ryu:2006bv,Hubeny:2007xt,Faulkner:2013ana,engelhardt2015quantum}.
Tensor networks on hyperbolic graphs that have such error correcting properties \cite{pastawski2015holographic,hayden2016holographic} give calculable models of AdS/CFT and demonstrate many expected holographic features of quantum gravity. 
However the networks are mostly discrete and it is not obvious that
they can be used beyond just models of AdS/CFT.  In particular we would like to work with a continuum QFT and understand how the holographic error correcting structure arrises in this language.
Algebraic QFT \cite{haag2012local} is the most concrete approach to studying the local quantum information aspects of QFT \cite{Witten:2018lha}, and so it is natural to seek the holographic error correcting structure in this approach.

In this paper we aim to place holographic error correcting codes on a firmer footing, by translating and generalizing these results fully to the language of infinite dimensional von Neumann algebras.
This program was initiated by Kang-Kolchmeyer \cite{Kang:2018xqy} and we wish to further expand on that work. 
Here we give a first pass at this problem, by studying exact error correction. There are several known shortcomings to such an exact
approach \cite{Kelly:2016edc,Hayden:2018khn,Akers:2019wxj}. However we find the structure at the exact level is quite natural, and we expect that at least some of it survives approximation. 
In a forthcoming work we will discuss approximate holographic error correcting codes where this expectation seems to be borne out. 
Also some of our results bear at least a passing resemblance to the new paper \cite{Akers:2020pmf}, which utilizes approximate error correction.

A recent hint for how to proceed came from the paper \cite{Casini:2019kex} studying superselection sectors \cite{Doplicher:1971wk} and entanglement in QFT. It was speculated that holographic CFTs should be thought of as containing a sub-theory -- low energy semi-classical gravity a.k.a. the bulk theory -- that comes along with a large number of charge sectors and associated local intertwiners arising from the UV quantum gravity degrees of freedom. The area term in the black hole entropy then can be understood as arising from the intertwiners, and in particular the author's argued that this perspective gives a robust understanding of how the area term secretly knows about the microscopic/UV degrees of freedom of quantum gravity, yet is still part of the bulk theory.
 
In this paper we will strengthen the connection between the ideas presented in \cite{Casini:2019kex} and the error correction approach to holography. 
We will start from the assumption of complementary recovery \cite{Harlow:2016vwg} and derive some of the expectations about the holographic map discussed in \cite{Casini:2019kex}. 
We will however rarely use the terms
``superselection-sector'' and ``intertwiner'' and instead highlight the importance of the existence of a consistent assignment of conditional expectations to boundary regions from
which many of our conclusions are drawn.  For example, we use this consistent set of conditional expectations  to show that the 
assumption of exact complementary recovery 
leads to the
same mathematical structure as that of QFT superselection-sectors as  formalized with a ``standard net of inclusions'' \cite{Longo:1994xe}.

A (non-commutative) conditional expectation is a mapping\footnote{It is completely positive, normal (that is ultra-weakly continuous) and unital. In this paper we work in $\infty$ dimensions so it is important to keep track of relevant continuity requirements. See for example \cite{ohya2004quantum}. } between a von Neumann algebra and a sub-algebra $E : \mathcal{M} \rightarrow \widetilde{\mathcal{N}}$ for some $\widetilde{\mathcal{N}} \subset \mathcal{M}$
that satisfies $E(mn) = E(m) n$ where $m\in \mathcal{M}$ and $n \in \widetilde{\mathcal{N}}$. A prominent example includes a normalized partial trace for a matrix
algebra $\mathbb{M}_A = \mathcal{B}(\mathcal{H}_A)$ acting on a finite dimensional Hilbert space $\mathcal{H}_A = \mathcal{H}_{BC}$ with the subalgebra $1_C \otimes \mathbb{M}_B
= 1_C \otimes \mathcal{B}(\mathcal{H}_B)$:
\begin{equation}
E(\cdot) = \frac{1_C}{{\rm dim} \mathcal{H}_C} \otimes {\rm Tr}_C(\cdot)
\end{equation}
Another important example is a group average over a symmetry:
\begin{equation}
\label{eavg}
E(\cdot) =  \int D g U(g) ( \cdot ) U(g)^{\dagger}
\end{equation}
where the subalgebra is the fixed point algebra of $U(g)$. 

In this first pass at modeling AdS/CFT, complementary recovery will lead to $E$'s mapping each local boundary algebra to a bulk algebra, and this map is consistent under restriction to subalgebras.
The Schr\"{o}dinger version of such a map
(for the global algebra) maps bulk states to boundary states and is often taken as the holographic map.  The program of entanglement wedge reconstruction is the attempt to reverse this map locally.  
We do not have an explicit conjecture for the form of these conditional expectations in AdS/CFT\footnote{A natural conjecture, which comes from the standard AdS/CFT lore, is that
the conditional expectation restricts to a sector of the CFT that is described by low dimension operators and large-$N$ factorization. This sector is approximately described by generalized free fields.
 All such conjectures involve a large parameter $N$ that roughly counts the degrees of freedom of the CFT. In this paper this parameter should be thought of as controlling the
 area law entropies that scale with $N^2$. For more precise formulations of what $E$ can be in this context see \cite{Casini:2019kex}.
 }, instead we will demonstrate the existence of $E$ based on general quantum information arguments and the expected behavior of entanglement in AdS/CFT as elucidated by the RT formula and generalizations \cite{Jafferis:2015del,Dong:2016eik,Dong:2017xht}.  

The conditional expectation in turn leads directly back to a version of the RT area formula. States that are left invariant under $E $
are factorized via the tensor product decomposition involving the relative commutant $\mathcal{M} \cong  \widetilde{\mathcal{N}} \otimes (\mathcal{M} \wedge \widetilde{\mathcal{N}}')$.
This particular tensor decomposition applies for type-I sub-factors, but a similar decomposition arises for general von Neumann algebras and this is a  consequence of Takasaki's theorem \cite{takesaki1972conditional} for $E$.
It is easy to show that applying the conditional expectation to operators in the relative commutant must result in a scalar multiple of the identity, or more generally it must give an operator in the center $Z( \widetilde{\mathcal{N}})\ni \pi_a$:
\begin{equation}
E(n^c) = \sum_a \pi_a \chi_a(n^c) \,, \qquad n^c \in \mathcal{N}^c =\mathcal{M} \wedge \widetilde{\mathcal{N}}'
\end{equation}
where $\chi_a$ is a family of normal states. These are exactly the code states in the language of \cite{Dong:2019piw} - that is the states that determine properties of the code itself and not properties of the specific states in the code subspace. The natural area operator is:
\begin{equation}
\mathcal{L}_{\widetilde{\mathcal{N}}} = \sum_a S(\chi_a) \pi_a
\end{equation}
where $ S(\chi_a)$ are the von Neumann entropies of these states. This operator arrises in the tensor decomposition of a state satisfying $\rho = \rho \circ E$ where $\rho \in \mathcal{M}_\star$.
That is:
\begin{equation}
S(\left. \rho \right|_{\mathcal{N}^c}) = \sum_a \rho(\pi_a) S(\chi_a)
= \rho(\mathcal{L}_{\widetilde{\mathcal{N}}})
\end{equation}

It is tempting to speculate that  an average such as \eqref{eavg} is at play and is related to the other averages that are discussed in the context of low dimensional holographic theories \cite{Saad:2019lba,Marolf:2020xie}
and in random tensor network models of holography \cite{Hayden:2016cfa}. 
The average in \eqref{eavg} is also the appropriate form of the conditional expectation for studying charged superselection sectors in QFT. In this case $U$ is related to a local symmetry
such that \eqref{eavg} acts locally and can thus be restricted consistently to local algebras. 
A Haar average, as in \eqref{eavg} arises also in the random tensor network models of holography where however now the average is taken over tensors that are determined by random local unitaries
on a bulk hyperbolic like network.

Despite this success we will also highlight some important differences between the structure of exact complementary recovery, and holographic CFTs. 
In particular we consider a setup discussed in \cite{Kelly:2016edc} involving overlapping boundary regions. This setup can be used to show that bulk algebras in our code are additive for overlapping boundary regions. This is certainly not true in holographic theories. Such additivity is well known from the theory of superselection-sectors \cite{Dong:2019piw}. The argument for additivity hinges on the assumption of the existence of a cyclic and separating vector through the code for the algebra generated by the overlapping boundary regions. This assumption is mild for infinite dimensional algebras, and indeed if the CFT vacuum is in the code-subspace it must be true.
So the resolution, as discussed in \cite{Kelly:2016edc}, is to move beyond exact recovery where cyclic and separating vectors lose their power.

We would now like to highlight several further results that we obtain in this paper:
\begin{itemize}
\item A clarification of the role of the Petz map in the original holographic code of \cite{Harlow:2016vwg}. In particular we highlight  a strong connection to the paper by Accardi-Cecchini \cite{accardi1982conditional}. This paper originally studied what is now known as the Petz map. As a precursor to this they defined a certain dual map that turns out to underpin most of the results in  \cite{Harlow:2016vwg}. In \cite{accardi1982conditional} the Petz map was called a generalized conditional expectation and they worked out the
 condition under which the generalized conditional expectation would turn into a regular conditional expectation and indeed this is the same mechanism at play in our paper. (Theorem~\ref{thm1}-\ref{thm2}).
\item A loosening of the assumptions on cyclic and separating states in \cite{Kang:2018xqy} that generalized \cite{Harlow:2016vwg} to purely infinite von Neumann algebras. (Theorem~\ref{thm1})
And a connection between Takesaki's theorem \cite{takesaki1972conditional} and the new paper of \cite{Gesteau:2020rtg}. Via this Theorem, the structure introduced in \cite{Gesteau:2020rtg} is exactly that of a conditional expectation. In particular the results of \cite{Gesteau:2020rtg} can be understood as the converse of some of our results. (Remark~2.\ref{rem:tak})
\item A connection between the so called Jones basic construction, that arises in index theory \cite{Jones:1983kv,Longo:1989tt}, and the structure that arises from complementary recovery. (Remark~2.\ref{rem:jones})

\item Some steps towards understanding the boundary dual of HRT/black hole area entropy in AdS/CFT. In particular we show the importance of entropies on the relative commutant associated to the conditional expectation. (Section~\ref{sec:area}). 

em An RT like minimization procedure that can be used to construct the conditional expectation. (Theorem~\ref{thm:min}).

\item An entanglement wedge nesting theorem for bulk algebras, that is well studied in holography. (Theorem~\ref{ewn}.) 

\item A new condition on holographic codes, called dual additive, that we argue can lead to phase transitions for the complementary reconstructable algebras as the boundary regions are varied.
We study these codes by introducing a new subspace called the split code. 
We use this and a version of the quantum minimality discussed in Theorem~\ref{thm:min} to prove a holographic entropy bound.

\end{itemize}

In this paper we will mostly work in $\infty$ dimensions. In which case our algebras (all von Neumann algebras) will  either be composed of type-$I_\infty$ algebras (referred to as the type-$I$ setting) or composed of type-$III_1$ algebras (referred to as the purely infinite setting), the later of which is appropriate for local algebras of QFT.  This dichotomy will allow us to discuss the main case of interest (QFT algebras), but at the same time discuss results that require finite entropies. These entropies, when associated to local algebras, should be thought of as regularized in some way that results in such a type-$I$ setting, but for the most part we will not explore a precise way to do this in this paper. The type-I setting is also appropriate for the algebras associated to the thermofield double state of a QFT, which is also a very important setting for discussing black hole entropies. Central decompositions will always be assumed to be discrete/atomic.  Our Hilbert spaces will all be separable and von Neumann algebras are always $\sigma$-finite. We will make some further assumptions about the type of relative commutants throughout the paper.

\subsection{Notation and conventions}

We spell out some notation and conventions here. Consider a Hilbert space $\mathscr{H}$ with a vector $\psi$. 
We define $\omega_{\psi}$ as the positive linear functional induced by the vector $\left< \psi \right| \cdot \left| \psi \right>$ defined on bounded operators of the Hilbert space $\mathcal{B}(\mathscr{H})$. 
Consider a von Neumann algebra $\mathcal{A}$ acting on a Hilbert space $\mathscr{H}$. 
A general state $\rho$ of $\mathcal{A}$ is an ultra-weakly continuous positive (normal) linear functional that is normalized $\rho(1) = 1$. 
It is an element of the predual $\rho \in \mathcal{A}_\star$. 
We define the support of a state as the smallest projection $\pi \equiv \pi(\rho) \in \mathcal{A}$ that satisfies $\rho(\pi) = 1$. Faithful states have $\pi=1$.
The support of a normalized vector $\psi$ will be defined as $\pi_{\mathcal{A}}(\psi) \equiv  \pi( \omega_{\psi}|_{\mathcal{A}} )$ and similarly for the commutant $\pi_{\mathcal{A}'}(\psi) \equiv  \pi( \omega_{\psi}|_{\mathcal{A}'} )$. Cyclic (separating) vectors for $\mathcal{A}$ have $\pi_{\mathcal{A}'}(\psi) = 1$ ($\pi_{\mathcal{A}}(\psi) = 1$ ).
We will define a \emph{quantum channel} $\gamma$ as a normal completely positive linear map between von Neumann algebras $\gamma: \mathcal{A} \rightarrow \mathcal{B}$. That is we will mostly work in the Heisenberg picture.
A normal *-homomorphism is a quantum channel that satisfies $\gamma(a_1 a_2) = \gamma(a_1) \gamma(a_2)$. 
We will refer to a normal injective unital *-homomorphism simply as an embedding. 
We can define the adjoint of a quantum channel $\gamma$ as an action on states $\gamma^\dagger(\rho) \equiv \rho \circ \gamma$ (which is in the Schr\"{o}dinger picture).
The set of normal faithful conditional expectations for $\mathcal{B} \subset \mathcal{A}$ will be denoted $C(\mathcal{A},\mathcal{B})$. 

\section{Operator algebra error correction}

In this section we will formulate the main theorem governing the error correcting code in \cite{Harlow:2016vwg} using general $\sigma$-finite von Neumann algebras along the lines of \cite{Kang:2018xqy}. 
We state the theorems here with slightly different assumptions and we give a proof (in Appendix~\ref{app:proof}) that follows the approach in \cite{accardi1982conditional}. The theorem is slightly more general than \cite{Kang:2018xqy}
since we remove the assumptions on the existence of cyclic and separating vectors through the code.  This is then more consistent with the finite dimensional results \cite{Harlow:2016vwg}. 

The error correcting code is determined by an isometry $V$ mapping:
\begin{equation}
V : \mathscr{H} =  \mathscr{H}_{\rm code} \rightarrow \mathscr{K}=  \mathscr{H}_{\rm phys}
\end{equation}
with $V^\dagger V = 1_{\mathscr{H}}$. On these respective Hilbert spaces we assume that there is a von Nuemann algebra $\mathcal{N} \subset \mathcal{B}(\mathscr{H})$ and  $\mathcal{M} \subset \mathcal{B}(\mathscr{K})$. We will not assume any specific relation between these from the outset.

\begin{theorem}[Operator algebra error correction]
\label{thm1} 
Using the above definitions the following statements are equivalent [assuming there is some vector $\psi \in \mathscr{H}$ such that $V\psi$ is cyclic and separating for $\mathcal{M}$] :
\begin{itemize}
\item[(i)] For any two vectors $\psi,\phi \in \mathscr{H}$:
\begin{equation}
\label{channel} 
\left. \omega_\psi \right|_{\mathcal{N}'} =\left. \omega_\phi \right|_{\mathcal{N}'}  \implies \left. \omega_{V\psi} \right|_{\mathcal{M}'}  = \left. \omega_{V\phi} \right|_{\mathcal{M}'} 
\end{equation}
\item[(ii)] The map $\alpha'(m') = V^\dagger m' V$ is a [faithful] unital quantum channel satisfying:
\begin{equation}
\alpha' : \mathcal{M}' \rightarrow \mathcal{N}' \qquad 
\end{equation}
\end{itemize}
\begin{itemize}
\item[(iii)] There exists a 
normal injective [unital] *-homomorphism:
\begin{equation}
\beta : \mathcal{N} \rightarrow \mathcal{M}
\end{equation}
with the property that for all  
$n \in \mathcal{N}$:
\begin{align}
\label{chan}
\beta(n) V &= V n
\end{align}
\end{itemize}
where either all bracketed statements [ $\ldots$ ] are included in the statement of the theorem or non of them are.

\end{theorem}

\begin{remark}

Some miscellaneous comments:

\begin{enumerate}
\item[(a)] Aside from appearing in the AdS/CFT literature, the (ii) $\leftrightarrow$ (iii) part of this Theorem is well known, and was proven long ago \cite{arveson1969subalgebras,accardi1982conditional} in various forms and using different approaches. It can also be found more recently in the quantum information literature \cite{crann2016private} as a correspondence between private and correctable sub-algebras.\footnote{The later paper also gives an approximate version which uses an approach based on Kretschmann et al.'s continuity theorem for Stinespring dilations \cite{kretschmann2008continuity} that has also recently appeared in the holographic context \cite{Hayden:2018khn}.}
For this reason we relegate the proof to Appendix~\ref{app:proof}.

\item[(b)] The physical interpretation of Theorem~\ref{thm1} of these results is hopefully clear. The condition on the linear functionals (i) is the sometimes called the ``DHW'' condition and was the main input from AdS/CFT that was used in \cite{Dong:2016eik} to prove bulk reconstruction.  Note that this condition is often equivalently stated in terms of the vanishing of relative entropies. It implies that no information, encoded in the differences in the state for $\mathcal{N}$, is carried to the environment $\mathcal{M}'$. 
Condition (ii) is essentially the same but stated in the Heisenberg picture, and (iii) is the statement of entanglement wedge reconstruction. In AdS/CFT bulk operators $n$ may be reconstructed on the boundary $\beta(n)$ and agree when acting on the code subspace $V$.

\item[(c)] The map $\beta$ is unique in the case where there is some $V\psi$ which is cyclic and separating for $\mathcal{M}$.

\item[(d)] While we have exact recovery as expressed through (iii) above there is a sense in which we have not actually started with a quantum channel and reversed it. Common conditions for the ability to reverse a quantum channel involve certain information constraints on states through the code, and we have not imposed any such constraint. 
For example it might seem more natural to start with the channel applied to $\mathcal{M}$:
\begin{equation}
\alpha : \mathcal{M} \rightarrow \mathcal{B}(\mathscr{H}) \,\qquad  \alpha (m) \equiv V^\dagger m V
\end{equation}
and try to construct a $\beta$ that reverses this, at least for the sub-algebra $\mathcal{N}$. 
Indeed such an approach has been successfully implemented in AdS/CFT by appealing to the JLMS condition on equality of relative entropies through the code \cite{Cotler:2017erl}. 
If we assume the existence of a factorized states $\sigma = \sigma_{\mathcal{N}} \otimes  \sigma_{\mathcal{N}'}$ then the Schr\"{o}dinger version of the quantum channel $\alpha^\dagger: \mathcal{B}(\mathscr{H})_\star \rightarrow \mathcal{M}_\star$ can be reduced to a quantum
channel for states in $\mathcal{N}_\star$ via:
\begin{equation}
\alpha_{\mathcal{N}}^\dagger : \mathcal{N}_\star  \rightarrow \mathcal{M}_\star \,, \qquad \alpha^\dagger_{\mathcal{N}}(\rho)
= \alpha^\dagger ( \rho \otimes \sigma_{\mathcal{N}'} )
\end{equation}
The JLMS condition:
\begin{equation}
\label{JLMS}
S_{\rm rel}\left( \left. \varrho_1 \right|_{\mathcal{N}} | \left. \varrho_2 \right|_{\mathcal{N}} \right) = 
S_{\rm rel}\left( \left. \alpha^\dagger \varrho_1 \right|_{\mathcal{M}} | \left.  \alpha^\dagger  \varrho_2 \right|_{\mathcal{M}} \right)  \,, \qquad \varrho_1,\varrho_2 \in \mathcal{B}(\mathscr{H})_\star
\end{equation}
upon setting $\varrho_{1,2} = \rho_{1,2} \otimes  \sigma_{\mathcal{N}'}$ becomes $S_{\rm rel}( \rho_1 | \rho_2 ) = S_{\rm rel}( \alpha_{\mathcal{N}}^\dagger \rho_1 | \alpha_{\mathcal{N}}^\dagger  \rho_2 )$.
This then proves the channel can be reversed using the Petz map \cite{Petz:1986tvy}. 
 In the ``DHW'' approach we will eventually derive the stronger JLMS condition from \emph{complementary recovery}, but we do not get it from Theorem~\ref{thm1}.  
Note that the approach based on \eqref{JLMS} is not easily generalizable to infinite dimensions because of the assumed existence of
a factorized state.\footnote{Actually one can write $\alpha^\dagger_{\mathcal{N}}
= \alpha^\dagger \circ P^\dagger$ where $P: \mathcal{B}(\mathscr{H}) \rightarrow \mathcal{N}$
is a conditional expectation. The adjoint is also known as a state extension. Such a conditional expectation is not guaranteed to exists for general von Neumann
algebras, and is not directly related to the conditional expectations that we discuss in this paper.}

\item[(e)] It is expected \cite{Dong:2017xht} that the DHW condition (i) is correct in AdS/CFT to all orders in a $G_N$ expansion, at least
if one focuses on the so-called reconstructable wedge \cite{Akers:2019wxj,Hayden:2018khn} and not the entanglement wedge. For small code subspaces these are generically
not expected to be much different.  
The JLMS condition however is known to receive perturbative corrections at higher orders in $G_N$. We will assume exact recovery here,
although even non-perturbative corrections can do a number on these results, as we will discuss in Section~\ref{sec:tensions}.
\end{enumerate}
\end{remark}

So far we have not used all of the power of complementary recovery which we turn to now.

\section{Complementary recovery}

Let us introduce some jargon. 

\begin{definition}[Reconstructable] We say that a bulk algebra $\mathcal{N}$ is reconstructable from a boundary algebra $\mathcal{M}$ if:
\begin{equation}
\omega_{\psi}|_{\mathcal{N}'} = \omega_{\phi}|_{\mathcal{N}'}  \quad  \rightarrow  \quad\omega_{V \psi}|_{\mathcal{M}'} = \omega_{V \phi}|_{\mathcal{M}'}\,, \qquad \forall \psi \in \mathscr{H}
\end{equation} 
We say that $\mathcal{N}$ is standardly reconstructable from $\mathcal{M}$ if in addition there is
a cyclic and separating vector $V \psi$ for $\mathcal{M}$. 

We further say that $\mathcal{N}$ is c-reconstructable from $\mathcal{M}$ if  $\mathcal{N}$ is reconstructable from $\mathcal{M}$ 
and  $\mathcal{N}'$ is reconstructable from $\mathcal{M}'$. The definition of standardly c-reconstructable follows the same pattern.
\end{definition}

\begin{remark}
Note that if $\mathcal{N}_1$ and $\mathcal{N}_2$ are both reconstructable from $\mathcal{M}$ then so is $\mathcal{N}_1 \vee \mathcal{N}_2$ (since by Theorem~\ref{thm1} $\alpha' : \mathcal{M}' \rightarrow \mathcal{N}_1' \wedge \mathcal{N}_2'$.) Also for any $\mathcal{N}_1 \subset \mathcal{N}$, with $\mathcal{N}$ reconstructable from $\mathcal{M}$ then $\mathcal{N}_1$ must be reconstructable from $\mathcal{M}$. Thus this condition is rather weak. c-reconstructability is a much stronger condition.
\end{remark}

A strong motivation for this condition comes from AdS/CFT and in particular its success at giving rise to the RT like formula in \cite{Harlow:2016vwg} as we will
see shortly. 
The notion of c-reconstructability relates two bulk ``regions'' for $\mathcal{N}$ and $\mathcal{N}'$ which we are implicitly assuming
have algebras associated to them that are commutants of each other. In the motivational case of the gravitational limit of AdS/CFT this need not be the case, a phenomena sometimes
referred to as the break down of complementary recovery. 
In particular if our code subspace it too large/permissive it is known that this assumption fails even for a single connected boundary region \cite{Hayden:2018khn,Akers:2019wxj}. Thus our discussion should be understood
as applying to/modeling AdS/CFT with relatively small code subspaces.
Given the structure we outline below we think this assumption should be compared to that of ``Haag duality'' in Algebraic QFT. 

If we have a bulk algebra $\mathcal{N}$, standardly c-reconstructable from $\mathcal{M}$,  Theorem~\ref{thm1} applied twice gives the embeddings $\beta, \beta'$ and quantum channels
$\alpha', \alpha$ which are both given by appropriate restrictions of $V^\dagger( \cdot )V$. We also get von Neumann subalgebras defined as $ \beta(\mathcal{N})\equiv \mathcal{N}^\beta \subset \mathcal{M}$ and  $\beta'(\mathcal{N}') 
\equiv (\mathcal{N}')^\beta \subset \mathcal{M}'$. 
Finally we get conditional expectations:

\begin{theorem}
\label{thm2}
If $\mathcal{N}$ is standardly c-reconstructable from $\mathcal{M}$, then the following properties hold:
\begin{itemize}
\item[(a)] $E = \beta \circ \alpha : \mathcal{M} \rightarrow \mathcal{N}^\beta$ is a faithful normal conditional expectation, $E \in C(\mathcal{M},\mathcal{N}^\beta)$. 
\item[(b)]  For any two normal states $\rho_{1,2}
\in \mathcal{N}_\star$:
\begin{equation}
\label{eqsrel}
S_{\rm rel}( \rho_1 | \rho_2) = S_{\rm rel}( \rho_1 \circ \alpha | \rho_2 \circ \alpha) 
\end{equation}
\item[(c)] \label{rem:tak} The bulk and boundary modular operators, $(J_{\mathcal{N}}, \Delta_{\mathcal{N}})$ and $(J_{\mathcal{M}},\Delta_{\mathcal{M}})$ for the respective vectors $\left| \eta \right>$ and $V\left| \eta \right>$ are related via:
\begin{align}
\label{modflow}
V \Delta_{\mathcal{N}}^{it} V^\dagger  &= e  \Delta_{\mathcal{M}}^{it} \,\,\, \implies V \Delta_{\mathcal{N}}^{it}  = \Delta_{\mathcal{M}}^{it}V    \\
V J_{\mathcal{N}}  V^\dagger &= eJ_{\mathcal{M}} \,\,\, \implies V J_{\mathcal{N}} = J_{\mathcal{M}} V   \\
 \beta( \sigma_{\mathcal{N}}^t( n) )&= \sigma_{\mathcal{M}}^t( \beta(n))\,,\qquad  \beta( j_{\mathcal{N}}( n) )= j_{\mathcal{M}}( \beta(n))\,, \qquad \forall \, n \in \mathcal{N}  
 \end{align}
\end{itemize}
And the same properties hold for the commutants, via the replacement
$(\alpha,\beta,\mathcal{N},\mathcal{M},E) \rightarrow(\alpha',\beta',\mathcal{N}',\mathcal{M}',E')$.

\end{theorem}
Note that we use standard definitions of the modular operators for a vector, see for example \cite{Ceyhan:2018zfg} for these definitions and conventions.
The modular automorphism group is $\sigma^{t}_{\mathcal{M}}(m) \equiv \Delta_{\mathcal{M}}^{it} m  \Delta_{\mathcal{M}}^{-it}$
and $j_{\mathcal{M}}(m) \equiv J_{\mathcal{M}} m J_{\mathcal{M}}$ and similar definitions for $\mathcal{N}$.

\begin{proof}
\noindent (a) Firstly note that $ \alpha \circ \beta (n) = n$.
Then $E$ satisfies $E(\beta(n)) = \beta \circ \alpha \circ \beta(n) = \beta(n)$, so indeed it fixes the sub-algebra $\mathcal{N}^\beta$. 
We can also explicitly calculate:
\begin{align}
\label{enm}
E( \beta(n) m ) &=
\beta ( V^\dagger \beta(n) m V )  = \beta (V^\dagger V V^\dagger  \beta(n)  m  V ) \\ & =  \beta (V^\dagger   \beta(n)  V V^\dagger m  V ) = \beta( \alpha(\beta(n)) \alpha(m)  )
= \beta(n) \beta(\alpha(m))  = 
\beta(n) E(m) 
\end{align}
which is the defining feature of a conditional expectation.

\noindent (b) We use monotonicity of relative entropy twice:
\begin{equation}
S_{\rm rel}( \rho_1 | \rho_2 ) \geq S_{\rm rel}( \rho_1 \circ \alpha | \rho_2 \circ \alpha ) \geq 
S_{\rm rel}( \rho_1 \circ \alpha \circ \beta | \rho_2 \circ \alpha \circ \beta )  = S_{\rm rel}( \rho_1 | \rho_2 ) 
\end{equation}
(Note that if we only demand that $\mathcal{N}$ is standardly reconstructable from $\mathcal{M}$ then the only inequality that we do not have is the first inequality since there is no quantum
channel from $\mathcal{M} \rightarrow \mathcal{N}$.) 

\noindent (c) These results are well known. They follow simply from Takeaski's theorem \cite{takesaki1972conditional}, which guarantees that for a state fixed by the conditional expectation $\omega_{V\eta}$ the action of the modular group on the fixed point algebra is stable:  $\sigma^{t}_{\mathcal{M}}(\mathcal{N}^\beta)  \subset \mathcal{N}^\beta$.  Uniqueness of the modular automorphism group \cite{stratila1981modular}, as a one parameter family of automorphisms that  fixes the state $\omega_{V\eta}$ and satisfies the KMS condition, then gives the equality of flows for the sub-algebra $\mathcal{N}^\beta$.  This becomes $\Delta_{\mathcal{M}}^{it} e = \Delta_{\mathcal{N}^\beta}^{it}$, $J_{\mathcal{M}}e = J_{\mathcal{N}^\beta}$. 
Note the isomorphism between $\mathcal{N}$ and $\mathcal{N}^\beta$ guarantees $ J_{\mathcal{N}^\beta} = V J_{\mathcal{N}} V^\dagger$ and $ \Delta_{\mathcal{N}^\beta} = V \Delta_{\mathcal{N}} V^\dagger$.

\end{proof}

\begin{remark}
\label{rem2}
There are several other properties that are obvious or well known, but it is useful to record these here:
\begin{enumerate}
\item $E$ fixes the code subspace $\omega_{V \eta} \circ E = \omega_{V \eta}$ for all $\eta \in \mathscr{H}$.
\item The code subspace projector satisfies $e \equiv VV^\dagger \in (\mathcal{N}^\beta)' \vee ((\mathcal{N}')^\beta)'$ and implements the conditional expectation:
\begin{equation}
E(m) e = e m e
\end{equation}
\item  Part (c) goes under the slogan ``bulk modular flow = boundary modular flow''. See for example \cite{Jafferis:2015del,Faulkner:2017vdd}. They were derived in the infinite dimensional case in \cite{Kang:2018xqy}. 

\item While we have stated the result for standard c-reconstructions, there should be a similar result for the non-standard case. At this point however we would have to confront
the fact that the von Neumann sub-algebra $\mathcal{N}^\beta$ does not have the same unit as $\mathcal{M}$ (it really acts on a different Hilbert space $\beta(1) \mathscr{K}$.) The conditional expectation, while still satisfying \eqref{enm}, is no longer unital and this is somewhat non-standard and less treated in the literature. Thus we leave this to future work. 

\item The support projectors associated to $\eta \in \mathscr{H}$ 
in the general (not necessarily standard) case satisfy:
\begin{align}
\pi_{(\mathcal{(N')}^\beta)'} = V \pi_{\mathcal{N}} V^\dagger  \leq \pi_{\mathcal{M}} \leq  \beta( \pi_{\mathcal{N}}) = \pi_{\mathcal{N}^\beta}  \\
  \pi_{(\mathcal{N}^\beta)'} = V \pi_{\mathcal{N}'} V^\dagger \leq \pi_{\mathcal{M}'} \leq \beta'( \pi_{\mathcal{N}'})  = \pi_{(\mathcal{N}')^\beta}
\end{align}
These follow from: $V \mathcal{N}' \left| \eta \right> \subset \mathcal{M}' V \left| \eta \right> $, $\omega_{V\eta}(\beta(\pi_\mathcal{N})) = 1$ and $\beta(\pi_\mathcal{N})^2 = \beta(\pi_\mathcal{N})$ and the fact that the isomorphism (the injective homomorphism) preserves support projectors. In the standard case  (with $V\eta$ is cyclic and separating for $\mathcal{M}$) then all support projectors are unit except:  $\pi_{(\mathcal{N}^\beta)'}= \pi_{(\mathcal{(N')}^\beta)'}  = e$.

\item \label{rem:jones} In the standard case there is also a connection to Jones' basic construction \cite{Jones:1983kv}. The idea is to introduce a new von Neumann algebra $\mathcal{M}_1$ that forms a chain of inclusions $  \mathcal{N}^\beta \subset \mathcal{M} \subset \mathcal{M}_1 $
defined as:
\begin{equation}
\mathcal{M}_1 = \mathcal{M} \vee e  = J_{\mathcal{M}} (\mathcal{N}^\beta)' J_{\mathcal{M}}
\end{equation}
 Equality of these two algebras follows from (i) $\mathcal{M} \vee  e=J_{\mathcal{M}} (\mathcal{M}' \vee e)  J_{\mathcal{M}}$ and (ii) $\mathcal{N}^\beta = \mathcal{M} \wedge \{ e \}'$. (i)
is a result of $\left[e, J_{\mathcal{M} } \right] =0$ which derives from \eqref{modflow} and (ii) follows
since all $x \in \mathcal{M}$ that commute with $e$ must satisfy $x\in \mathcal{N}^\beta$. This in turn follows from the separating property of $ V \eta$ for $\mathcal{M}$ since $ x V\left| \eta \right> =e x V\left| \eta \right> = E(x) V\left| \eta \right>$ so $x = E(x) \in \mathcal{N}^\beta$. 

But it is now easy to show that:
\begin{equation}
\beta(n)  V \left| \eta \right>
= J_{\mathcal{M}}^2 V n \left| \eta \right> = J_{\mathcal{M}} V J_{\mathcal{N}} n \left| \eta \right>
= J_{\mathcal{M}} V  J_{\mathcal{N}} n J_{\mathcal{N}} \left| \eta \right>
= J_{\mathcal{M}}  \beta'( J_{\mathcal{N}} n J_{\mathcal{N}} ) J_{\mathcal{M}}  V \left| \eta \right>
\end{equation}
so that the separating property for $\mathcal{M}$ gives $\beta(n) =  J_{\mathcal{M}}  \beta'( J_{\mathcal{N}} n J_{\mathcal{N}} ) J_{\mathcal{M}} $ and thus:
\begin{equation}
J_{\mathcal{M}} \mathcal{N}^\beta  J_{\mathcal{M}} =  (\mathcal{N}')^\beta = \mathcal{M}_1'
\end{equation}
 We learn that the basic construction is related to the complementary bulk region, and in particular it's failure to be self-dual on the physical Hilbert space,
 that is $ \mathcal{M}_1= ((  \mathcal{N}')^\beta )' \neq \mathcal{N}^\beta$. The minimal Jones index, should it exist, roughly measures how many times $\mathcal{N}^\beta$ fits inside $\mathcal{M}$, and
 the Jones index of the inclusion $\mathcal{M}_1 \subset \mathcal{M}$ is the same. 

\item The structure of complementary recovery can now be given a standard quantum error correction interpretation. We have a noisy channel which is $\alpha$
and the recovery channel $\beta$. 
If we were to to choose to work exclusively on the Hilbert space $\mathcal{K}$ then the noisy channel should be thought of as the conditional expectation $E:   \mathcal{M} \rightarrow \mathcal{N}^\beta$
and the recovery map is the inclusion $\iota : \mathcal{N}^\beta \rightarrow \mathcal{M}$.  We also have $E' : \mathcal{M}' \rightarrow (\mathcal{N}')^\beta$ and  $\iota' : (\mathcal{N}')^\beta \rightarrow \mathcal{M}'$.

\end{enumerate}
\end{remark}

We turn to some consequences of this structure.

\section{Area law}
\label{sec:area}

We now show how to derive the so called Black Hole/RT area law discussed in \cite{Harlow:2016vwg} from this mathematical structure. In finite dimensions we will effectively re-derive the result of \cite{Harlow:2016vwg}.
Beyond this we would like to consider two situations. Firstly we would like to allow for infinite dimensional type-$I_\infty$ algebras and in this case our results are new. 
This would for example apply to the classic thermofield double/wormhole duality of \cite{Maldacena:2001kr} where the bulk algebras are associated to the two exterior regions of the two black holes.

Secondly we would like to make some more speculative  comments  in the purely infinite setting. This is relevant to bulk algebras that end on the boundary of AdS
where the associated boundary algebras are known to be composed of type-III von Neumann algebras \cite{haag2012local}.   
The bulk and boundary entropies are infinite now. However there should be a useful notion of generalized entropy in quantum gravity that is finite deep in the bulk. Since the remaining infinity comes from an IR divergence of the area near the boundary of AdS, in principle we should be able to factor this out. One might expect that the entropy formula then becomes more like the type-$I_{\infty}$ setting. 
As a first step towards this we show that with some extra assumptions an area operator can still be usefully defined in this setting.

\subsection{Type-$I$ setting}

\label{sec:t1}

We assume that $\mathcal{N}$ is standardly c-reconstructable from $\mathcal{M}$ and these algebras are composed of only type-$I$ algebras (possibly of infinite dimensions.) 
Our ultimate goal is to compute the von Neumann entropy $S_{\mathcal{M}}(\rho \circ \alpha)$ of some state $\rho \in \mathcal{N}_\star$ from the code subspace.
We could use a vector state $\rho = \omega_{\eta}|_{\mathcal{N}}$ where $\eta \in \mathscr{H}$ but this is not a necessary restriction.
Since it is in the code subspace we have invariance under the conditional expectation: $\rho \circ \alpha = \rho \circ \alpha \circ E$. 

We will assume that the center of $\mathcal{M}$ is trivial $Z(\mathcal{M}) \equiv \mathcal{M} \wedge \mathcal{M}' = \mathbb{C} 1$. If we had not made this assumption we should be able to re-derive and generalize the results in \cite{Kamal:2019skn}. There are fairly good reasons however to assume that the boundary algebras (at least for simple boundary regions such as intervals or spheres) in a CFT have trivial center \cite{guido1996conformal}. The more general situation also becomes a mess of indices that we don't feel the need to treat at this point.  

The inclusion $\mathcal{N}^\beta \subset \mathcal{M}$ decomposes under the center $Z(\mathcal{N}^\beta)$
as:
\begin{equation}
\label{inca}
\mathcal{N}^\beta = \bigoplus_a \mathcal{N}^\beta_a \, \qquad \mathcal{N}^\beta_a \equiv \pi_a \mathcal{N}^\beta  \subset \pi_a \mathcal{M} \pi_a \equiv \mathcal{M}_a
\end{equation}
where $\pi_a$ are minimal projectors in $Z(\mathcal{N}^\beta)$ satisfying $\sum_a \pi_a = 1$, and these new von Neumann algebras act on $\pi_a \mathscr{K}$. 
We have assumed the center has a discrete decomposition. 
This decomposition also applies to the bulk algebra:
\begin{equation}
\label{centN}
\mathcal{N} = \bigoplus_a \mathcal{N}_a \,, \qquad \mathcal{N}_a = \varpi_a \mathcal{N} \,, \qquad \varpi_a = \alpha(\pi_a)
\end{equation}
where $\varpi_a$ are a complete set of central projections for $\mathcal{N}$.
We show that $\varpi_a$ are in one to one correspondence to $\pi_a$,
as follows. The form $\varpi_a=\alpha(\pi_a)$ is still a projection $\varpi_a^2 = V^\dagger \pi_a e \pi_a V = V^\dagger \pi_a^2 V = \varpi_a$ that is since $\pi_a$ commutes with $e = VV^\dagger \in (\mathcal{N}^\beta)'$.
Similarly $\varpi_a n = V^\dagger \pi_a \beta(n) V = n \varpi_a$ such that $\varpi_a$ commutes with $\mathcal{N}$ and is thus in $Z(\mathcal{N})$. Also any $\varpi_a$ that is in $Z(\mathcal{N})$ gives a projection $\beta(\varpi_a)$ that commutes with $\mathcal{N}^\beta$ and so is in $ Z(\mathcal{N}^\beta)$. 

The conditional expectation satisfies: $E( \pi_a m) = \pi_a E(m) = \pi_a E(m) \pi_a = E( \pi_a m \pi_a)$ so there are associated normalized conditional expectations:
\begin{equation}
E_a : \mathcal{M}_a \rightarrow \mathcal{N}_a^\beta \,, \qquad E_a( m_a) \equiv  E (\pi_a m_a \pi_a) = \pi_a E(m_a)
\end{equation}
which are faithful since $E$ is faithful. That is $E_a \in C(\mathcal{M}_a,\mathcal{N}_a^\beta)$. Then:
\begin{equation}
E(m) =\sum_a 
E(\pi_a m) 
=  \sum_a 
E_a(\pi_a m \pi_a) 
\end{equation}
and this decomposition extends to the action on all of $\mathcal{M}$, which does not have a central decomposition.
The state of interest, being invariant under $E$, decomposes after inserting the resolution of the identity:
\begin{equation}
\label{summ}
\rho\circ \alpha (m) = \sum_a \rho \circ \alpha (m \pi_a) =  \sum_a \rho \circ \alpha (\pi_a m \pi_a) = \sum_a p_a \rho_a \circ \alpha(m_a)
\end{equation}
where $m_a = \pi_a m \pi_a \in \mathcal{M}_a$ and  $\rho_a( \cdot) \equiv \rho(\varpi_a \cdot)/p_a$ and $p_a = \rho(\varpi_a)$.  So that $\rho_a \circ \alpha \in (\mathcal{M}_a)_\star$.

Define the relative commutant:
\begin{equation}
\mathcal{N}^c_a \equiv (\mathcal{N}_a^\beta)' \wedge \mathcal{M}_a
\end{equation}
We have the chain of inclusions: $\mathcal{N}^\beta_a  \subset \mathcal{N}^\beta_a \vee \mathcal{N}^c_a   \subset  \mathcal{M}_a $.
The later inclusion has commutative relative commutant $( \mathcal{N}^\beta_a \vee \mathcal{N}^c_a)^c = Z(\mathcal{N}^c_a )
$. Furthermore in the type-I setting  one has $ Z(\mathcal{N}^c_a) = Z(\mathcal{N}_a^\beta \vee \mathcal{M}_a') = Z(\mathcal{N}_a^\beta) \vee Z( \mathcal{M}_a)= \mathbb{C} 1_a$, since $\mathcal{N}_a^\beta \subset \mathcal{M}_a$
 is automatically split so $\mathcal{N}_a^\beta \vee \mathcal{M}_a' \cong \mathcal{N}^\beta \otimes \mathcal{M}_a'$. 
This then implies equality:
\begin{equation}
\label{conormal}
\mathcal{M}_a = \mathcal{N}^\beta_a \vee \mathcal{N}^c_a 
\end{equation}
and this will be one of the simplifications for the type-I case. More generally an inclusion which satisfies \eqref{conormal} is called conormal in $\mathcal{M}_a$ \cite{longo1984solution}.

Takesaki proved generally \cite{takesaki1972conditional} that the existence of $E_a$ implies the factorization result:
\begin{equation}
\label{nbnc}
\mathcal{N}^\beta_a \vee \mathcal{N}^c_a \cong  \mathcal{N}^\beta_a \otimes \mathcal{N}^c_a
\end{equation}
where a state on the code subspace $\rho \in \mathcal{N}_\star$ factorizes: 
\begin{equation}
\rho_a \circ \alpha (\beta(n_a) n^c_a) = \rho_a \circ \alpha \circ E_a (\beta(n_a) n^c_a)
= 
\rho_a(n_a)\,\, \rho_a \circ \alpha \circ E_a (n^c_a)
\end{equation}
for all $n_a \in \mathcal{N}_a$ and $n^c_a \in \mathcal{N}^c_a$. 
Now it is not hard to see that $E_a (n^c_a)$ commutes with $\mathcal{N}^\beta$ so it must be in the center. 
That is $E_a(n^c_a) = \pi_a \chi_a(n^c_a)$ for some normalized faithful normal state  $\chi_a$ on $\mathcal{N}^c_a$. 
Thus the final form of the factorized state is:
\begin{equation}
\label{facte}
\rho_a \circ \alpha (\beta(n_a) n^c_a) = \rho_a(n_a) \chi_a(n^c_a)
\end{equation}

We can compute the entropy using the decomposition in \eqref{summ}:
\begin{equation}
S_{\mathcal{M}}( \rho \circ \alpha ) = \sum_a   p_a( S_{\mathcal{M}_a}(\rho_a \circ \alpha) - \ln p_a)
\end{equation}
and using the factorization property:
\begin{equation}
S_{\mathcal{M}}( \rho \circ \alpha ) = \sum_a   p_a( S_{\mathcal{N}_a}(\rho_a)  + S_{\mathcal{N}^c_a}(\chi_a) - \ln p_a)
= S_{\mathcal{N}}(\rho) + \sum_a p_a S_{\mathcal{N}^c_a}(\chi_a)
\end{equation}
where $S_{\mathcal{N}}$ is the usual von Neumann entropy for a state with a central decomposition.
Defining the ``area operator'':
\begin{equation}
\mathcal{L}_{\mathcal{N}} = \sum_a \varpi_a S_{\mathcal{N}^c_a}(\chi_a)
\end{equation}
gives the final form:
\begin{equation}
S_{\mathcal{M}}( \rho \circ \alpha ) 
=  \rho(\mathcal{L}_\mathcal{N})+S_{\mathcal{N}}(\rho)  
\end{equation}
We will also write this in the following form, which can be more convenient in various context, as an entropy
on the physical Hilbert space:
\begin{equation}
S_{\mathcal{M}}( \varrho ) = S_{\mathcal{N}^\beta}( \varrho|_{\mathcal{N}^\beta} ) + \varrho( \mathcal{L}_{\mathcal{N}^\beta})
\equiv S_{\rm gen}^{\mathcal{N}^\beta}(\varrho) \,, \qquad \varrho = \rho \circ \alpha
\end{equation}
where:
\begin{equation}
\mathcal{L}_{\mathcal{N}^\beta} = \sum_a \pi_a S_{\mathcal{N}^c_a}(\chi_a)
\end{equation}
This might be more convenient since we can consider $S_{\rm gen}^{\mathcal{N}^\beta}(\varrho)$
when $\varrho \in \mathcal{M}_\star$ is not on the code subspace. Although then it certaintly does not equal $S_\mathcal{M}(\varrho)$.

\subsection{Comments on the purely infinite setting}

\label{sec:area-inf}

If $\mathcal{M}$ is a localized boundary algebra then on general grounds it will contain type-III factors.
This excludes the possibility of defining entropies. However from our experience with AdS/CFT and entanglement in gravity we expect there is 
a meaningful way to compute entropies (really generalized entropies) at least in the bulk while assuming $G_N$ is finite but small.
Since the previous cases can be thought of as studying the black hole area of a thermofield double, we don't expect that in the bulk the entropies for boundary anchored regions are all that different from the above discussion.
The main difference is that now the bulk region ends on the boundary. Any UV issues with defining von Neumann entropy should come from there, so we expect that the bulk algebra $\mathcal{N}$ will still be purely infinite but the divergences associated to entanglement entropy have a different origin - correlations near the boundary and not correlations near the RT surface.  
Since entanglement is non-local it is not easy to make this intuition precise, however the area operator is local, so based on this we expect that there should be still a useful notion of area even in the purely infinite case. Divergences in the area from the boundary are related to UV issues, but we can isolate these by considering difference in the expectation of the area operator in two states that only differ locally in the bulk. The picture we expect to emerge is that the area operator becomes an unbounded operator affiliated with the center. For now it seems that
the best option for the unboundedness is from the central sum itself. 

Indeed we can define an operator $\mathcal{L}_{\mathcal{N}}$ quite generally, with a few seemingly reasonable assumptions.
 We still have a relative commutant $\mathcal{M} \wedge (\mathcal{N}^\beta)'$ which also has a central decomposition. We assume that $\mathcal{M} \wedge (\mathcal{N}^\beta)'$ is made up of a discrete sum of type-I factors. 
 Then $E$ gives rise to the normal states $\chi_a$ on the relative commutant as usual. It is possible that there are divergences from the entropy of a fixed $\chi_a$ being infinite (as there could be above!) when the central decomposition contains type-$I_\infty$ factors, however this seems to go against the locality of the area law divergence in AdS/CFT. Rather we assume the real divergences come from the central sum. 
Thus we retain the area operator:
\begin{equation}
\label{newarea}
\mathcal{L}_{\mathcal{N}} = \sum_a \varpi_a S_{\mathcal{N}^c_a}(\chi_a)
\end{equation}
and this can certainly be unbounded if $S_{\mathcal{N}^c_a}(\chi_a)$ are finite but not uniformly bounded for all $a$. 
Roughly speaking we might expect the central elements represent different parts of the RT surface and those elements approaching the boundary have diverging $S_{\mathcal{N}^c_a}(\chi_a)$. 

The utility of this definition remains to be seen. One issue is that in the purely infinite case there can be other kinds of inclusions that don't factorize across the relative commutant.
That is the inclusion $\mathcal{N}^\beta \subset \mathcal{M}$ may not be conormal. An extreme example of this,  in the type-II or type-III setting, is the possibility to have non-trivial inclusions of factors (with no center) that yet has a trivial relative commutant.  Such an inclusion is called irreducible or singular.
As mentioned the area operator \eqref{newarea} is still well defined, and still derives from a factorized state associated to \eqref{nbnc}.
However the factorized state lives on  $\mathcal{N}^\beta_a \vee \mathcal{N}^c_a$
where the inclusion $\mathcal{N}^\beta_a \vee \mathcal{N}^c_a \subset \mathcal{M}_a$ is now non-trivial. There is however a unique conditional expectation:\footnote{It is unique because there is a bijection between the space of conditional expectations $C( \mathcal{M}_a, \mathcal{N}^\beta_a \vee \mathcal{N}^c_a)$ and $C(( \mathcal{N}^\beta_a \vee \mathcal{N}^c_a)^c,Z( \mathcal{N}^\beta_a \vee \mathcal{N}^c_a))$ which is now trivial \cite{baillet1988indice,Longo:1989tt}. }
\begin{equation}
E_a^2 : \mathcal{M}_a \rightarrow \mathcal{N}^\beta_a \vee \mathcal{N}^c_a
\end{equation}
which factorizes the original conditional expectation:
\begin{equation}
E_a = E_a^{1} E_a^{2}\,,  \qquad E_a^1 : \mathcal{N}^\beta_a \vee \mathcal{N}^c_a \rightarrow \mathcal{N}^\beta_a 
\end{equation}
The area operator is then naturally associated to $E_a^{1}$ and what we do with $E_a^2$ is not clear to us. 
It is interesting to speculate that index theory will play an important role \cite{Jones:1983kv,kosaki1986extension,Longo:1989tt}, since this gives a natural way to assign an entropy to $E_a^2$ \cite{pimsner1986entropy}. 

Another possible issue with this interpretation is the following. The complementary recovery condition for multiple boundary and bulk regions turns out to be intimately related to a formal abstraction of the theory of superselection sectors,
as we will discuss further in Section~\ref{sec:gen}. However based on this structure and certain reasonable sounding assumptions about the conformal symmetries of the boundary theory and how it acts on the bulk algebras, one can argue that the algebras $\mathcal{N}$ (as well as those of $\mathcal{M}$) have trivial center which would be a problem for defining a useful notion of an area operator. We expect the fact that there are many more issues with exact complementary recovery applied in this more general setting, suggests we should not take this problem too seriously. 

\subsection{Maximally mixed codes}

There is a distinguished conditional expectation that we would like to highlight in this section. 
These are associated to the maximally mixed codes that were shown to be important for AdS/CFT \cite{Akers:2018fow,Dong:2019piw,Dong:2018seb}. 
We assume that the dimension of the relative commutants $\mathcal{N}_a^c $ are all finite, although the central sum could be infinite. 
In other words we demand that $\mathcal{N}_a^c$ is a factor of type-$I_{d_a}$ for some integers $d_a \geq 1$. 
The argument proceeds by noting that the computation
of Renyi entropies in AdS/CFT should be thought of as a computation directly in the code subspace, since one uses the Euclidean path integrals and effective field theory. 
For example if we work in a type-I setting, then one can define a density matrix $\mathcal{D}_{\mathcal{M}}$ affiliated with $\mathcal{M}$ and such that $\omega_{V\psi}|_{\mathcal{M}}(\cdot)  = {\rm Tr}_{\mathcal{M}}( \mathcal{D}_{\mathcal{M}} (\cdot) )$
where ${\rm Tr}_{\mathcal{M}}$ is the unique normal semi-finite tracial weight on $\mathcal{M}$. The easiest way to impose the Renyi entropy constraint is to demand that
 $(\mathcal{D}_{\mathcal{M}})^{is} V \left| \psi \right> = V \left| \xi_s \right>$ for some vector $\xi_s$ where $s\in \mathbb{R}$.\footnote{Since $\mathcal{D}_{\mathcal{M}}$ can be unbounded the imaginary power $s\in\mathbb{R}$ behaves better than an integer power $n \rightarrow is$. However if we assume that $(\mathcal{D}_{\mathcal{M}})^{n/2}$  is in the domain of $V \left| \psi \right>$ for all integers $n \geq 0$, the former constraint
 is implied by the condition that $(\mathcal{D}_{\mathcal{M}})^{n/2} V \left| \psi \right>$ is in the code subspace by Carlson's theorem. This later condition then connects to the
 Renyi entropy computations since the norm of this vector computes the $n+1$'th Renyi entropy. }

Since the state is factorized \eqref{facte} we must have $\mathcal{D}_{\mathcal{M}} = \sum_{a} \pi_a \mathcal{D}_{{\mathcal{N}_a^c} } \otimes \mathcal{D}_{\mathcal{N}^\beta_a}$. 
Now consider the linear functional  $\omega_{V\psi,(\mathcal{D}_{\mathcal{M}})^{is} V \psi } (\cdot) \equiv \left<\psi\right| V^\dagger (\cdot) (\mathcal{D}_{\mathcal{M}})^{is} V\left| \psi \right> $. 
Applied to the relative commutant $n^c_a \in \mathcal{N}^c_a$ we have:
\begin{equation}
\omega_{V\psi,(\mathcal{D}_{\mathcal{M}})^{is} V \psi } (n^c_a)= \omega_{V\psi,(\mathcal{D}_{\mathcal{M}})^{is} V \psi } (\pi_a) \chi_a^{(s)}(n^c_a)\,,
\qquad \chi_a^{(s)}(n^c_a) = {\rm Tr}_{\mathcal{N}_a^c} (\mathcal{D}_{{\mathcal{N}_a^c}}^{1+is} n^c_a)
\end{equation}
If this was on the code it would have the form:
\begin{equation}
\omega_{V\psi, V\xi_s}(n^c_a) = \omega_{\psi,\xi_s}(\alpha(n^c_a)) = \omega_{\psi,\xi_s}( \alpha \circ E(n^c_a)) =
\omega_{V\psi,V\xi_s}( \pi_a) \chi_a(n^c_a)
\end{equation}
So for equality we must at least require that $D_{\mathcal{N}^c_a} \propto 1_{\mathcal{N}^c_a}$, which also requires that the relative commutants for fixed $a$ are finite 
dimensional. In fact these conditions are sufficient, as can be seen by computing $\omega_{V\psi,(\mathcal{D}_{\mathcal{M}})^{is} V \psi } \circ E(m) =
\left< \psi \right| V^\dagger m V V^\dagger ( \mathcal{D}_{\mathcal{M}})^{is} V \left| \psi \right> = \omega_{V\psi,(\mathcal{D}_{\mathcal{M}})^{is} V \psi }(m)  $ where we used the fact that the maximally mixed condition implies that $(\mathcal{D}_{\mathcal{M}})^{is} \in \mathcal{N}^\beta $
and so it commutes with $VV^\dagger$. 

We can give a simple characterization of the conditional expectation that leads to such a code. $E$ is tracial when restricted to the relative commutant.
That is:
\begin{equation}
E(n_c^1 n_c^2) = E(n_c^2 n_c^1)   \,, \qquad n_c^{1,2} \in \mathcal{N}^c 
\end{equation}
and note that while we motivated this in the type-I setting it is also a constraint we could consider imposing on the properly infinite case.
Indeed this is one of the condition that is necessary for a conditional expectation to give rise to the minimal Jones/Kosaki/Longo index \cite{Jones:1983kv,kosaki1986extension,Longo:1989tt}.

\begin{definition}
We say that $\mathcal{N}$ is standardly c-reconstructable from $\mathcal{M}$ with a maximally mixed code if the resulting conditional expectation
is tracial on the relative commutant. 
\end{definition}

\section{Quantum minimality}
\label{sec:nm}

One way to find an algebra $\mathcal{N}$ that is standardly c-reconstructable from $\mathcal{M}$ is simply to minimize the ``generalized entropies'' of all reconstructable algebras.
We will establish this with two theorems of increasing level of precision, yet decreasing generality. We start with:

\begin{theorem}[Quantum minimality]
\label{thm:min}
In the type-I setting, given an algebra $\mathcal{N}$ that is standardly reconstructable from $\mathcal{M}$ then for all $\psi \in \mathscr{H}$
such that $S_{\mathcal{M}}(\omega_{V\psi}) < \infty$ we have the estimate:
\begin{equation}
\label{sgen}
S_{\mathcal{M}}(\omega_{V\psi}) \leq  \sum_a S_{(\mathcal{M} \wedge (\mathcal{N}^\beta)')_a}(\omega_{V\psi}) + S_{\mathcal{N}}(\omega_{\psi})
\end{equation}
where the sum is over the central decomposition of $\mathcal{N}^\beta$.  

Equality is achieved in \eqref{sgen}, for a fixed $\psi$ assumed to be cyclic for $\mathcal{N}$, iff $\mathcal{N}$ is standardly c-reconstructable from $\mathcal{M}$.
\end{theorem}
\begin{proof}
The forward implication is clear from Section~\ref{sec:t1}.
The converse is a  consequence of positivity of mutual information for the state $V\psi$.
Define the decoherence conditional expectation:
\begin{equation}
D(m) = \sum_a \pi_a m \pi_a
\end{equation}
with fixed point algebra $\bigoplus_a \mathcal{M}_a$. This preserves the trace on $\mathcal{M}$ (which exists in the type-$I$ setting) by cyclicity and completeness
$\sum \pi_a = 1$.  It is then well known (see for example \cite{Casini:2019kex}) that the difference in entropies is a relative entropy:
\begin{equation}
-S_{\mathcal{M}}(\omega_{V\psi} \circ D) + S_{\mathcal{M}}(\omega_{V\psi}) = - S_{\rm rel}(\omega_{V\psi}|\omega_{V\psi} \circ D)
\end{equation}
In which case:
\begin{align}
\label{betin}
S_{\mathcal{M}}(\omega_{V\psi})& -  S_{\mathcal{N}^\beta}(\omega_{V\psi})- \sum_a S_{(\mathcal{M} \wedge (\mathcal{N}^\beta)')_a}(\omega_{V\psi}) \\
&= -  S_{\rm rel}(\omega_{V\psi}|\omega_{V\psi} \circ D) -  \sum_a I( \mathcal{N}^\beta_a : (\mathcal{M} \wedge (\mathcal{N}^\beta)')_a) \leq 0
\end{align}
where we used the fact that:
\begin{equation}
\omega_{V\psi} \circ D (m)=  \sum_a p_a (\omega_{V\psi})_a(m)
\end{equation}
and we used the definitions below \eqref{summ} and also:
\begin{equation}
I( \mathcal{N}^\beta_a : (\mathcal{M} \wedge (\mathcal{N}^\beta)')_a) =  S_{(\mathcal{M} \wedge (\mathcal{N}^\beta)')_a}( (\omega_{V\psi})_a)
+  S_{\mathcal{N}^\beta_a}( (\omega_{V\psi})_a)-  S_{\mathcal{M}_a}( (\omega_{V\psi})_a)
\end{equation}
We have the *-isomorphism $\mathcal{N} \cong \mathcal{N}^\beta$, derived from the injective *-homomorphism. Under this isomorphism the following states are clearly equivilent:
$\omega_{V\psi}|_{\mathcal{N}^\beta}$ and $ \omega_{V\psi} \circ \beta |_{\mathcal{N}}$. But note that  $\omega_{V\psi} \circ \beta(n)
= \omega_{\psi}(n)$ for $n \in \mathcal{N}$.  Thus the entropies agree $S_{\mathcal{N}^\beta}(\omega_{V\psi})= S_{\mathcal{N}}(\omega_{\psi})$. 
In particular we do not need c-recovery to conclude this. Then \eqref{betin} becomes to \eqref{sgen}.

We have equality in \eqref{betin} if the state $\omega_{V\psi}$ on $\mathcal{M}$ is a block diagonal sums of states on $\mathcal{M}_a$ and if
under this decomposition each state is factorized across the $a$-relative commutants. 
This implies that there is a 
conditional expectation $E_\psi : \mathcal{M} \rightarrow \mathcal{N}^\beta$ that preserves
this state:\footnote{We do not need the explicit form of $E_\psi$ but for completeness we give it in this footnote:
\begin{equation}
E_{\psi}(m) =\bigoplus_a  1_{\mathcal{N}^c_a} \left( {\rm Tr}_{\mathcal{N}^c_a} \otimes {\rm Id}_{\mathcal{N}^\beta} \right)(\mathcal{D}_{\mathcal{N}^c_a} \pi_a m \pi_a)\,,\qquad m \in \mathcal{M}
\end{equation}
where $ {\rm Tr}_{\mathcal{N}^c_a}(\cdot) \otimes {\rm Id}_{\mathcal{N}_a^\beta}$ is the partial trace on the relative commutant $\mathcal{N}_a^c$.
We think of a partial trace here as an (nsf) operator valued weight $\in P(\mathcal{N}^c_a \otimes {\mathcal{N}_a^\beta}, {\mathcal{N}_a^\beta})$ where
this can be derived by extending the tracial weight, which is a special case of an operator valued weight $ {\rm Tr}_{\mathcal{N}^c_a} \in  P(\mathcal{N}^c_a, \mathbb{C})$, to act on a larger algebra.
Also $D_{\mathcal{N}^c_a}$ is the density matrix affiliated with $\mathcal{N}^c_a$ defined such that
$\omega_{V\psi} |_{\mathcal{N}_a^c} ( \cdot ) = {\rm Tr}_{\mathcal{N}^c_a}( \mathcal{D}_{\mathcal{N}^c_a} (\cdot) )$.}
\begin{equation}
\label{after}
\omega_{V\psi} \circ E_\psi = \omega_{V\psi} 
\end{equation} 

The conditional expectation $E_\psi$ also preserves the following state:
\begin{align}
\label{simarg}
\omega_{Vx \psi} \circ E_\psi (\cdot) &= \omega_{ V\psi} ( \beta(x^\dagger) E_\psi( \cdot) \beta(x) )
= \omega_{ V\psi} \circ E_\psi( \beta(x^\dagger) \cdot \beta(x) )\\&= \omega_{V\psi}(\beta(x^\dagger) \cdot \beta(x))
= \omega_{V x \psi} (\cdot)
\end{align}
 for any $x \in \mathcal{N}$. By the cyclicity of $\psi$ with respect to $\mathcal{N}$ we find that $E_\psi = E$ preserves all states on the code subspace $V(\mathscr{H})$.
This implies that $E$ is faithful since we know (by the standardness of the reconstruction assumption) there is a vector $V\eta$ for which $\omega_{V\eta}|_{\mathcal{M}}$ is faithful.

We thus have equality of relative entropy through the code:
\begin{equation}
S_{\rm rel}(\omega_{V\psi} | \omega_{V\phi} ; \mathcal{N}^\beta)
= S_{\rm rel}(\omega_{V\psi} | \omega_{V\phi} ; \mathcal{M})
\end{equation}
for all $\psi,\phi \in \mathscr{H}$.
But this implies (since $\omega_{V \psi} \circ \beta(n) = \omega_\psi(n)$) that for all $\psi,\phi \in \mathscr{H}$
\begin{equation}
(\omega_{\psi} - \omega_{\phi})|_{\mathcal{N}} = 0 \quad \implies \quad S_{\rm rel}(\omega_{V\psi}| \omega_{V\phi}; \mathcal{M}) = 0 \quad \implies (\omega_{V\psi} - \omega_{V\phi})|_{\mathcal{M}} =0
\end{equation}
which then implies that $\mathcal{N}'$ is standardly reconstructable from $\mathcal{M}$  which combines with the original reconstructability to give the c-reconstruction statement. 
\end{proof}

We note that this theorem sets up something that looks like an analog of the RT minimization. In fact it even looks like the quantum Engelhardt-Wall version \cite{engelhardt2015quantum} which involves
an extremization over the generalized entropy. Here the two terms on the right hand side of \eqref{sgen} might be thought of as the area term, from the relative commutant, and the bulk entropy terms respectively.  There are several features of this result that however are not satisfying. Firstly, it is not fair to call the right hand side of \eqref{sgen} a generalized entropy since the entropy on
the relative commutant will in general not evaluate to the expectation value of an operator, as it does at the minimal point. That is the area
operator becomes a state dependent operator and this is probably not what we expect from gravity. 

Secondly, the result is way too general, due to the following argument.
Assuming there is some von Neumann algebra, call it $\mathcal{E}$ that is standardly c-reconstructable from $\mathcal{M}$.  $\mathcal{E}$ is then like the entanglement wedge.
Now any other subalgebra $\mathcal{N} \subset \mathcal{E}$ is standardly reconstructable from $\mathcal{M}$ and thus satisfies the inequality \eqref{sgen}.\footnote{We will
show below in Theorem~\ref{ewn} that all standardly reconstuctable algebras must be subsets of $\mathcal{E}$.\label{fn}}
However we don't expect the generalized entropy to always increase for arbitrary subregions of the entanglement wedge. Indeed along the entanglement horizon
the generalized entropy must never increase by quantum focusing \cite{bousso2016quantum}.  This issue is also related to the fact that we should really be \emph{extremizing} the generalized entropy.
These two issues in some sense cancel each other out and we thus conjecture that the terms in \eqref{sgen} can only be interpreted as a generalized entropy in certain circumstances (for certain subalgebras $\mathcal{N}$ of $\mathcal{E}$).

Thirdly the minimization over \eqref{sgen} is all or nothing. If one state achieves the minimum then so do all the states in the code subspace.
This is simply a feature of exact complementary recovery and not something we will attempt to fix in this current paper. 

We give the following attempt to fix the first problem for maximally mixed codes:
\begin{theorem}[Quantum minimality for maximally mixed codes]
\label{thm:min2}
In the type-I setting, define  $\mu_\mathcal{M}$ to be the space of algebras $\mathcal{N}$ that are standardly reconstructable from $\mathcal{M}$ and where the associated relative commutants $\mathcal{M}_a \wedge (\mathcal{N}_a^\beta)'$ 
(see \eqref{inca}) are finite dimensional type-$I_{d_a}$ algebras, then:
\begin{equation}
\label{sgen2}
S_{\mathcal{M}}(\omega_{V\psi}) \leq \inf_{\mathcal{N} \in \mu_{\mathcal{M}}} \left( \omega_{V \psi}( \mathcal{L}_{\mathcal{N}^{\beta}}) + S_{\mathcal{N}}(\omega_{\psi}) \right)
\end{equation}
where $ \mathcal{L}_{\mathcal{N}^\beta} = \sum_a  \pi_a \ln d_a$.  Assume there is some $\mathcal{N}_0 \in \mu_{\mathcal{M}}$ such that 
$\psi$ is cyclic for $\mathcal{N}_0$ then equality is achieved in \eqref{sgen2} for some $\mathcal{N} \in \mu_{\mathcal{M}}$ with $\mathcal{N}_0 \subset \mathcal{N}$ iff $\mathcal{N}$ is standardly c-reconstructable from $\mathcal{M}$ with a maximally mixed code.
\end{theorem}
\begin{proof}
Consider the central sum of the relative commutant $\mathcal{N}^c =
\bigoplus_a \mathcal{N}_a^c = \bigoplus_a \mathcal{M}_a \wedge (\mathcal{N}^\beta_a)'$
where $\pi_a \in Z(\mathcal{N}^\beta)$ are the central projectors and recall that
$\mathcal{M}_a = \pi_a \mathcal{M} \pi_a$. There exists a conditional expectation $E$ such that:\footnote{Again we do not need the explicit form but we give it here for completeness.
\begin{equation}
E(m) =\bigoplus_a  \frac{1_{\mathcal{N}^c_a}}{d_a} \left( {\rm Tr}_{\mathcal{N}^c_a} \otimes {\rm Id}_{\mathcal{N}^\beta} \right)( \pi_a m \pi_a)\,,\qquad m \in \mathcal{M}
\end{equation}
}
\begin{equation}
S_{\mathcal{M}}(\omega_{V\psi}) -  \omega_{V\psi}( \mathcal{L}_{\mathcal{N}^\beta}) - S_{\mathcal{N}}(\omega_{\psi})
= - S_{\rm rel}( \omega_{V\psi} | \omega_{V\psi} \circ E)  \leq 0
\end{equation}
Thus equality is achieved when $\omega_{V\psi} \circ E = \omega_{V\psi}$. 
The conditional expectation $E$ also preserves the following state:
\begin{align}
\omega_{Vx \psi} \circ E (\cdot) &= \omega_{ V\psi} ( \beta(x^\dagger) E( \cdot) \beta(x) )
= \omega_{ V\psi} \circ E( \beta(x^\dagger) \cdot \beta(x) )\\&= \omega_{V\psi}(\beta(x^\dagger) \cdot \beta(x))
= \omega_{V x \psi} (\cdot)
\end{align}
 for any $x \in \mathcal{N}$. 
 By the cyclicity of $\psi$ with respect to $\mathcal{N} _0 \subset \mathcal{N}$ we find that $E$ preserves all states on the code subspace $V(\mathscr{H})$.
This implies that $E$ is faithful since we know (by the standardness of the reconstruction assumption) there is a vector $V\eta$ for which $\omega_{V\eta}|_{\mathcal{M}}$ is faithful.
The rest of the argument proceeds as in the proof of Theorem~\ref{thm:min} after \eqref{simarg}. 
\end{proof}

Note that this is only a partial solution to the first problem since $\mathcal{L}_{\mathcal{N}^\beta}$ is not manifestly a bulk operator, which we would
expect of an area operator. Indeed 
one can use $V^\dagger \mathcal{L}_{\mathcal{N}^\beta} V$ which is a bulk operator however it is not guaranteed to be in the center of $\mathcal{N}$.
Note that the argument in the paragraph after \eqref{centN} does not apply since we do not have c-reconstructability for $\mathcal{N}$. 
Indeed the best we can do is show that, assuming the existence of $\mathcal{E}$ that is standardly c-reconstructable from $\mathcal{M}$ with a maximally mixed code,
then for any other $\mathcal{N}$ described in the theorem $V^\dagger \mathcal{L}_{\mathcal{N}^\beta} V \in \mathcal{E}\wedge \mathcal{N}'$.
So we again conjecture that $V^\dagger \mathcal{L}_{\mathcal{N}^\beta} V \in Z(\mathcal{N})$ only in some more limited situations and that this resolves the tension with
the second problem mentioned above. We will give one example where this occurs in Section~\ref{disjoint}.

\section{Generalities}
\label{sec:gen}

We now describe some generalizations when there are multiple algebras acting on the same code subspace. 

\begin{theorem}[Entanglement wedge nesting]
If $\mathcal{N}_1$
is 
reconstructable from $\mathcal{M}_1$ and $\mathcal{N}_2'$ is 
reconstructable from $\mathcal{M}_2'$
with $\mathcal{M}_1 \subset \mathcal{M}_2$ then $\mathcal{N}_1 \subset \mathcal{N}_2$. 
\label{ewn}
\end{theorem}
\begin{proof}
We have:
\begin{equation}
\beta_1(\mathcal{N}_1) \equiv \mathcal{N}_1^\beta \subset \mathcal{M}_1 \subset \mathcal{M}_2
\end{equation}
Recall that if we don't have a standard reconstruction $\mathcal{N}^\beta_1$ is still a subset of $\mathcal{M}_1$ (the main difference being that in order to interpret it as a von Neumann algebra it must act on $\beta(1) \mathscr{K}$.)
So we can apply $\alpha_2 : \mathcal{M}_2 \rightarrow \mathcal{N}_2$ to $\beta_1(\mathcal{N}_1)$. Indeed, since $\alpha_2 \circ \beta_1(n_1) = V^\dagger \beta_1(n_1) V = n_1$, we must have:
\begin{equation}
\mathcal{N}_1 \subset  \alpha_2(\mathcal{M}_2) \subset \mathcal{N}_2 
\end{equation}
as required. 
\end{proof}
This theorem is then analagous to the well known constraint on entanglement wedges in AdS/CFT \cite{Wall:2012uf}. Many interesting dynamical properties arise from nesting, 
such as the boundary QNEC \cite{Koeller:2015qmn,Balakrishnan:2017bjg}. 

A corollary is that if $\mathcal{M}_1 = \mathcal{M}_2$ then we still have $\mathcal{N}_1 \subset \mathcal{N}_2$.
Thus we cannot do better than complementary recovery as promised in footnote~\ref{fn}.  Note that $\beta_1 \circ \alpha_2$ only makes sense if $\mathcal{N}_1 = \mathcal{N}_2$ when it becomes a conditional expectation.

\begin{theorem}[Induced conditional expectation] If $\mathcal{N}_1$ is standardly 
c-reconstructable from $\mathcal{M}_1$
and $\mathcal{N}_2$ is standardly c-reconstructable from $\mathcal{M}_2$ with $\mathcal{M}_1 \subset \mathcal{M}_2$ then:
\begin{equation}
\left. \beta_2 \right|_{\mathcal{N}_1} = \beta_1\,, \qquad \left.  E_2\right|_{\mathcal{M}_1} = E_1
\end{equation}
\label{induce}
\end{theorem}
\begin{proof}
From Theorem~\ref{ewn} we have $\mathcal{N}_1 \subset \mathcal{N}_2$. Pick $V\left| \eta \right>$ cyclic and separating
for $\mathcal{M}_2$, so:
\begin{equation}
\beta_2(n_1) V \left| \eta \right>
= V n_1 \left| \eta \right> = \beta_1(n_1) V \left| \eta \right>
\end{equation}
Thus $\beta_2(n_1) = \beta_1(n_1)$, and since $\alpha_2|_{\mathcal{M}_1} = \alpha_1$ is clear:
\begin{align}
E_2(m_1) = \beta_2 \circ \alpha_2(m_1) = \beta_2 \circ \alpha_1(m_1) = \beta_1 \circ \alpha_1(m_1) = E_1(m_1)
\end{align}

\end{proof}

\begin{definition}[Net of complementary codes] 
A net of von Neumann algebras over a partially ordered index set $\mathcal{I}$ is an assignment $M : i \in \mathcal{I}\rightarrow \mathcal{M}_i$ of von Neumann algebras on a Hilbert space $\mathscr{K}$ which preserves the order relation $\mathcal{M}_i \subset \mathcal{M}_k$ if $i\leq k$. A net of complementary codes consists of two nets $N$ and $M$ over the same index set acting respectively on $\mathscr{H}$ and $\mathscr{K}$, with an isometry relating these $V(\mathscr{H}) \subset \mathscr{K}$, 
and such that for every $i \in \mathcal{I}$, $\mathcal{N}_i$ is c-reconstructable for $\mathcal{M}_i$. We write  $N \rightarrow M$ and sometimes refer to this simply as a code.
The net $M$ is called standard if there is a vector $\Omega \in \mathscr{K}$ which is cyclic and separating for every $\mathcal{M}_i$. The code $N \rightarrow M$ is called standard if $M$ is standard and $N$ is standard for the same vector $V \left| \eta_\Omega \right> = \left| \Omega \right>$. In this case all c-reconstructions are standard.
A directed net derives from a directed index set $\mathcal{I}$. 
\label{def:ncc}
\end{definition}

Note that Theorem~\ref{ewn} guarantees that the order relation on the net $N$ is consistent with the reconstruction condition.
That is we need not have assumed that $N$ is an ordered net (ordered under inclusion in the same way as $\mathcal{I}$)
- rather we could derive it using Theorem~\ref{ewn}.

\begin{theorem}
A standard directed net of complementary codes $N \rightarrow M$ is equivialent  to a Longo-Rehren standard net of inclusions $N^\beta \subset M$ where:
\begin{equation}
N^\beta = \{ \mathcal{N}_i^\beta : i \in \mathcal{I} \} \,, \qquad \mathcal{N}_i^\beta = \beta_i(\mathcal{N}_i)
\end{equation} 
with the consistent assignment of conditional expectations $i \rightarrow E_i = \beta_i \circ \alpha_i$. 
\end{theorem}
\begin{proof}
For the forward implication we just need to check the consistency condition of the conditional expectation. That is $E_i =  \left.E_k \right|_{\mathcal{M}_i}$ whenever $i \leq k$ and indeed this follows
from Theorem~\ref{induce}. 

The converse statement follows from the definition of the standard LR net $\tilde{N} \subset M$ given in \cite{Longo:1994xe}. We summarize the structure here.
In addition to two subnets $\widetilde{N} \subset M$ (consistent assignments of von Neumann sub-algebras from a partially ordered index set), 
there is a code subspace $ \mathscr{K}_0 \subset \mathscr{K}$ with a vector  $\Omega \in \mathscr{K}_0$ that is cyclic and separating for all $\mathcal{M}_i$ and such that
$\widetilde{\mathcal{N}}_i \Omega$ is dense on $\mathscr{K}_0$. There are conditional expectations that are consistent under
restriction as above and such that $\omega_{\Omega} \circ E_i = \omega_{\Omega}$. 

We now reproduce the structure of complementary recovery. Define the code subspace projector $e \mathscr{K} = \mathscr{K}_0$.
Note that $e = \pi_{\widetilde{\mathcal{N}}_i'}(\Omega) $ for all $i$ (by definition). Define:
\begin{equation}
\mathscr{H} = \mathscr{K}_0\,, \quad V = e|_{\mathscr{H}}
\end{equation}
so that $ e = VV^\dagger$ and $V^\dagger V = 1_{\mathscr{H}}$. Then define $\eta \in \mathscr{H}$ such that $V \eta = \Omega$ (these are really the same vectors, but it is convenient to label them this way to compare with the complementary recovery structure.) 
Then consider the *-homomorphism $\phi_i: \widetilde{\mathcal{N}}_i \rightarrow \mathcal{N}_i$ defined via $\phi_i(\tilde{n}_i) = \tilde{n}_i e$ where $\tilde{n}_i \in  \widetilde{\mathcal{N}}_i $. 
It is a *-isomorphism since if $\phi_i(\tilde{n}_i) =0$
then $0 = \tilde{n}_i e \left| \Omega\right> =   \tilde{n}_i \left| \Omega\right>$ which implies that $ \tilde{n}_i  = 0$ by the separating property for $\mathcal{M} \ni  \tilde{n}_i $.
Note that  $\mathcal{N}_i$ acts on $\mathscr{H}$. Define $\beta_i(n_i) = \phi_i^{-1}(n_i)$ for $n_i \in \mathcal{N}_i$. 

Now by definition:
\begin{equation}
\beta_i(\phi_i(\tilde{n}_i)) V 
= \tilde{n}_i e
=  \phi_i(\tilde{n}_i)
= V \phi_i(\tilde{n}_i)
\end{equation}
So that $\beta_i(n_i) V = V n_i$ for all $n_i \in \mathcal{N}_i$ and we learn from  Theorem~\ref{thm1} that $\mathcal{N}_i$ is standardly reconstructable from $\mathcal{M}_i$

By definition the conditional expectation satisfies $E_i \circ \beta_i= \beta_i$. Also using the fact that:
\begin{equation}
\omega_{V\eta} \circ E_i = \omega_{V \eta}|_{\mathcal{M}_i}
\end{equation}
and by a similar argument to \eqref{simarg} we also have invariance under:
\begin{equation}
\omega_{Vn_i \eta} \circ E_i = \omega_{V n_i \eta}|_{\mathcal{M}_i}
\end{equation}
so assuming the cyclicity property of $\eta$ we find $V^\dagger E_i (\cdot) V = V^\dagger (\cdot)|_{\mathcal{M}_i} V$. This however implies that $V^\dagger m_i V \in \mathcal{N}_i$ since
$E_i(m_i) \in \beta_i(\mathcal{N})$ and $V^\dagger \beta_i(n_i) V = n_i$ for all $n_i \in \mathcal{N}_i$. The cyclicity property follows by the density of $\widetilde{\mathcal{N}}_i \left| \Omega \right>$
in $V(\mathscr{H})$ which implies that $ \mathcal{N}_i \left| \eta \right>$ is dense on $\mathscr{H}$.
Then $V^\dagger m_i V \in \mathcal{N}_i$ implies by Theorem~\ref{thm1},  that $\mathcal{N}_i'$ is standardly reconstructable from $\mathcal{M}_i'$.
Together we get the complementary recovery statement.

\end{proof}
The LR net of inclusions has many nice properties. In infinite dimensions it is most conveniently studied using Longo's canonical endomorphism \cite{Longo:1989tt}. We have not applied this to the case at hand and
we hope to return to this in the future.

\section{Disjoint regions and phase transitions}
\label{disjoint}
 
In this section we consider the interesting case of two boundary algebras $\mathcal{M}_1$ and $\mathcal{M}_2$
that are causally disjoint $\mathcal{M}_1 \subset \mathcal{M}_2'$, and we assume that these boundary algebras satisfy the split property \cite{Doplicher:1984zz}. 
This models the algebra of two disjoint regions of the boundary theory. For example,  it is useful to have in mind a 2d CFT where $\mathcal{M}_i$ arise from the local operators in two disjoint intervals on a fixed time slice. 
It is well known that in holographic theories the entanglement wedge for the vacuum of such boundary algebras undergoes a phase transition \cite{Headrick:2010zt,Faulkner:2013yia} as a function of the cross ratio of the end points of the intervals. We would like to model this situation abstractly in terms of an error correcting code. It is expected that new conditions must be added in order to achieve such a phase transition \cite{hartman2013entanglement,Hartman:2014oaa} since they should only arise for theories with a large-$N$ limit and some sparsity condition on the spectrum of low lying operators that is often linked to strong coupling.  We will give a conjecture, that we call dual-additivity, as to a condition that is important for such holographic phase transitions. 

The split property implies the existence of an isomorphism $\Phi:\mathcal{M}_1 \otimes \mathcal{M}_2 \rightarrow \mathcal{M}_1 \vee \mathcal{M}_2$, implemented with a unitary $\mathcal{U}:
\mathscr{K} \rightarrow \mathscr{K} \otimes \mathscr{K}$ that maps:
\begin{equation}
\Phi( \mathcal{M}_1 \otimes \mathcal{M}_2 ) = \mathcal{U}^\dagger \left( \mathcal{M}_1 \otimes \mathcal{M}_2 \right) \mathcal{U} = \mathcal{M}_1 \vee \mathcal{M}_2
\end{equation}
We will assume that there exists a $\Omega \in \mathscr{K}$ that is cyclic and separating for $\mathcal{M}_{1,2}$ and $\mathcal{M}_1 \vee \mathcal{M}_2$
(thus the split inclusion $\mathcal{M}_1 \subset \mathcal{M}_2'$ is standard \cite{Doplicher:1984zz}). 
We can uniquely specify $\mathcal{U}$ by requiring that $\mathcal{U} \left| \Omega \right> \equiv \left| \xi \right>$ and $  \left| \Omega \right> \otimes \left| \Omega \right>$ are in the same natural cone.
Define the split vector as $\left| S \right> = \mathcal{U}^\dagger (  \left| \Omega \right> \otimes \left| \Omega \right>)$ then $\omega_S(m_1 m_2) = \omega_\Omega(m_1) \omega_\Omega(m_2)$. 
The split property, as stated above, requires properly infinite algebras.

Now consider a code subspace $\mathscr{H}$ and an isometry $V: \mathscr{H} \rightarrow \mathscr{K}$.  We take this code subspace to be such that
$\mathcal{N}_1 \subset \mathcal{B}(\mathscr{H})$ is c-reconstructable from $\mathcal{M}_1$ and $\mathcal{N}_2  \subset \mathcal{B}(\mathscr{H}) $ is c-reconstructable from $\mathcal{M}_2$. 
 This implies, by Theorem~\ref{ewn} that $\mathcal{N}_1$ and $\mathcal{N}_2$ are causally separated ($\mathcal{N}_1 \subset \mathcal{N}_2'$). 
We demand that $\left| \Omega \right> \equiv V \left| \eta_\Omega \right> \in V(\mathscr{H})$ so these reconstructions are standard.
We also demand that there is a bulk algebra $\mathcal{N}_{12}$ that is standardly c-reconstructable from $\mathcal{M}_1 \vee \mathcal{M}_2$.

As a first result we show:

\begin{lemma} With the assumptions stated above, the bulk algebras $\mathcal{N}_{1,2}$ satisfy the split property and the inclusion $\mathcal{N}_1 \subset \mathcal{N}_2'$ is a standard split inclusion. 
Hence there exists a $\mathcal{U}_{\mathscr{H}}:\mathscr{H} \rightarrow \mathscr{H} \otimes \mathscr{H}$ such that:
\begin{equation}
\label{uH}
\mathcal{N}_1 \vee \mathcal{N}_2 = \mathcal{U}_{\mathscr{H}}^\dagger ( \mathcal{N}_1 \otimes \mathcal{N}_2) \mathcal{U}_{\mathscr{H}}
\end{equation}
uniquely defined so that  $\left| \eta_\Omega \right> \otimes \left| \eta_\Omega \right> $ and $\mathcal{U}_{\mathscr{H}} \left| \eta_\Omega \right> $ are in the same natural cone. 
Furthermore the bulk and boundary splitting unitaries are related via:
\begin{equation}
\label{Thv}
 \mathcal{U} ^\dagger V \otimes V = (\Theta')^\dagger  V \mathcal{U}_{\mathscr{H}}^\dagger
\end{equation}
with the unitary operator $\Theta' \in (\mathcal{N}_1^\beta \vee \mathcal{N}_2^\beta)'$. 
\end{lemma}
\begin{proof}
We know that:
\begin{equation}
\beta_1(\mathcal{N}_1)\equiv \mathcal{N}_1^\beta  \subset \mathcal{M}_1 \,, \qquad
\beta_2(\mathcal{N}_2)\equiv  \mathcal{N}_2^\beta \subset \mathcal{M}_2 
\end{equation}
and since $\beta_{12}$ always agrees with $\beta_{1,2}$ under restriction:
\begin{equation}
\beta_{12}( \mathcal{N}_1 \vee \mathcal{N}_2 )
= \beta_{12}( \mathcal{N}_1) \vee \beta_{12}(\mathcal{N}_2 )
= \mathcal{N}_1^\beta \wedge \mathcal{N}_2^\beta
= \Phi( \mathcal{N}_1^\beta \otimes \mathcal{N}_2^\beta)
\end{equation}
Also since $\beta_{12}$ is a *-homomorphism we can invert it on the range:
\begin{equation}
\mathcal{N}_1 \vee \mathcal{N}_2  =\beta_{12}^{-1} \left( \Phi \left( \mathcal{N}_1^\beta \otimes \mathcal{N}_2^\beta \right) \right)
= \beta_{12}^{-1} \circ \Phi \circ (\beta_1 \otimes \beta_2)  \left[  \mathcal{N}_1 \otimes  \mathcal{N}_2  \right]
\equiv \Phi_{\mathscr{H}}( \mathcal{N}_1 \otimes  \mathcal{N}_2  )
\end{equation}
which is the desired split isomorphism. We know that $\left| \eta_\Omega \right>$ is cyclic and separating for $\mathcal{N}_1,\mathcal{N}_2$
such that this vector is cyclic for  $\mathcal{N}_1  \vee \mathcal{N}_2$.
$\left| \eta_\Omega \right>$ is cyclic and separating for $\mathcal{N}_{12} \supset \mathcal{N}_1  \vee \mathcal{N}_2$ and so it is separating for $\mathcal{N}_1  \vee \mathcal{N}_2$.
Thus the split inclusion is standard. 

Thus there exists a $\mathcal{U}_{\mathscr{H}}$ as in \eqref{uH} defined as above.
Note that the linear functionals induced by the vectors $V \mathcal{U}_{\mathscr{H}}^\dagger \left| \eta_{\Omega } \right>  \otimes  \left| \eta_{\Omega } \right>$
and $\left| S \right>$ agree when restricted to operators in $\mathcal{N}_1^\beta \vee \mathcal{N}_2^\beta$. Thus there exists a unitary $\Theta' \in (\mathcal{N}_1^\beta \vee \mathcal{N}_2^\beta)'$
with: 
\begin{equation}
\label{aTh}
\Theta' \left| S \right> = V \mathcal{U}_{\mathscr{H}}^\dagger \left| \eta_{\Omega } \right>  \otimes  \left| \eta_{\Omega } \right>
\end{equation}
It is not hard to verify that the two splitting unitaries satsify:
\begin{equation}
 V^\dagger \otimes V^\dagger \mathcal{U} ( \mathcal{N}_1^\beta \vee \mathcal{N}_2^\beta ) \mathcal{U}^\dagger
V \otimes V = \mathcal{U}_{\mathscr{H}} V^\dagger ( \mathcal{N}_1^\beta \vee \mathcal{N}_2^\beta )V \mathcal{U}_{\mathscr{H}}^\dagger
\end{equation}
so we must have the relation \eqref{Thv} with the same unitary in \eqref{aTh}.

\end{proof}

Now consider the possible bulk algebras associated to $\mathcal{M}_1 \vee \mathcal{M}_2 $.
We have already assumed there is such an algebra $\mathcal{N}_{12}$ which is reconstructable from $\mathcal{M}_1 \vee \mathcal{M}_2 $.
In some sense this assumption contains in it a form of Haag duality for the bulk algebra associated to these disjoint boundary algebras. That is following from this assumption we also have
 $ \mathcal{N}_{(12)'} \equiv (\mathcal{N}_{12})' $ which is standardly reconstructable from $(\mathcal{M}_1 \vee \mathcal{M}_2)' $.

We have established above in Theorem~\ref{ewn} that $\mathcal{N}_{1},\mathcal{N}_2 \subset \mathcal{N}_{12}$ and hence:
\begin{equation}
\label{addv}
\mathcal{N}_1 \vee \mathcal{N}_2 \subset \mathcal{N}_{12} \left( = (\mathcal{N}_{(12)'})'  \right)
\end{equation}
But there might not be equality. We can also write this inclusion for algebras on the physical Hilbert space. Since $\mathcal{N}_{1,2} \subset \mathcal{N}_{12}$ we know from Theorem~\ref{induce} that $\beta_1 = \beta_{12}|_{\mathcal{N}_1}$ and  $\beta_2 = \beta_{12}|_{\mathcal{N}_2}$ so we can apply $\beta_{12}$ to \eqref{addv}
\begin{equation}
(\mathcal{N}_1)^\beta \vee (\mathcal{N}_2)^\beta \subset (\mathcal{N}_{12})^\beta \subset \mathcal{M}_{12} = (\mathcal{M}_{(12)'})' 
\subset ((\mathcal{N}_{(12)'})^\beta)'
\end{equation}
These inclusions were already observed by \cite{Casini:2019kex}.

Note that the first inclusion \eqref{addv} is referred to as an additivity violation since the reconstructable algebra $\mathcal{N}_{12}$ is not locally generated.
A measure of additivity violation is given by computing the difference in mutual informations  \cite{Longo:2017mbg,Xu:2018fsv,Casini:2019kex}:
\begin{equation}
\label{mid}
I( \mathcal{M}_1,  \mathcal{M}_2)
- I( \mathcal{N}_1^\beta,  \mathcal{N}_2^\beta) = S_{\rm rel}(\rho | \rho \circ (E_1 \otimes E_2)^\Phi)
\end{equation}
where $\rho \in (\mathcal{M}_{12})_\star$ with a vector representer $\rho = \omega_\psi$ where
$\psi \equiv V \eta_\psi$ is on the code subspace. Also $E_i$ are the conditional expectations for each disconnected region $\mathcal{M}_i \rightarrow \mathcal{N}_i^\beta$ and the mutual information is defined with the relative entropy:
\begin{align}
I( \mathcal{M}_1,  \mathcal{M}_2) &= S_{\rm rel}(\omega_\psi | \omega_{S_\psi}; \mathcal{M}_1 \vee \mathcal{M}_2) \\
I( \mathcal{N}_1^\beta,  \mathcal{N}_2^\beta) &= S_{\rm rel}(\omega_\psi | \omega_{S_\psi}; \mathcal{N}_1^\beta \vee \mathcal{N}_2^\beta) =
 S_{\rm rel}(\omega_{\eta_\psi} | \omega_{S_{\eta_\psi}}; \mathcal{N}_1 \vee \mathcal{N}_2)  =  I( \mathcal{N}_1,  \mathcal{N}_2) 
\end{align}
where $S_\psi = \mathcal{U}^\dagger( \left| \psi \right> \otimes \left| \psi \right> )$ and
in the bulk $S_{\eta_\psi} = \mathcal{U}_{\mathscr{H}}^\dagger ( \left| \eta_\psi \right> \otimes  \left| \eta_\psi \right> )$. We have used $\omega_{S_\psi} \circ \beta_{12} = (\omega_{\eta_\psi} \otimes \omega_{\eta_\psi})
\circ \Phi^{-1}_{\mathscr{H}}$. 
In the type-I setting, using Section~\ref{sec:area}, we can easily compute the left hand side of \eqref{mid}:
\begin{equation}
\label{eqn}
S_{\rm rel}(\rho | \rho \circ (E_1 \otimes E_2)^\Phi)
= - S_{\mathcal{M}_{12}}(\rho)  + S_{\mathcal{N}_{1}^\beta \vee \mathcal{N}_2^\beta}(\rho)
+ \rho(\mathcal{L}_{\mathcal{N}_1^\beta} )   +  \rho(\mathcal{L}_{\mathcal{N}_2^\beta} )  
\end{equation}

Positivity of this quantity looks like a quantum minimality condition for the generalized entropy. Motivated by this we give the following interpretation of additivity violations.
The boundary split state is invariant under the conditional expectation:
\begin{equation}
\omega_{S_\psi} = \omega_{S_\psi} \circ   (E_1 \otimes E_2)^\Phi\,,
\qquad (E_1 \otimes E_2)^\Phi = \Phi \circ (E_1 \otimes E_2) \circ \Phi^{-1}
\end{equation}
where $\Phi : \mathcal{M}_1 \otimes \mathcal{M}_2 \rightarrow \mathcal{M}_1 \vee \mathcal{M}_2$ is
the split isomorphism. 
We could also write: 
\begin{equation}
\omega_{S_\psi} = (\Phi^{-1})^\dagger (\rho|_{\mathcal{M}_1} \otimes \rho|_{\mathcal{M}_2} )
\end{equation}
Since $\rho$ is in the code subspace: $\rho = \rho \circ E_{12}$ for the conditional expectation associated to the c-reconstructable algebra $\mathcal{N}_{12}$. So right hand side 
oif \eqref{mid} can be written as:
\begin{equation}
S_{\rm rel}(\rho \circ E_{12} | \rho \circ (E_1 \otimes E_2)^\Phi)
\end{equation}
so in a sense we are comparing two conditional expectations. The order parameter for additivity violation is the extent to which $\rho$ is
left invariant by $ (E_1 \otimes E_2)^\Phi$. 

We will now try to sharpen this statement. The idea is that $ (E_1 \otimes E_2)^\Phi$ should be associated to a new code subspace. The question of additivity violation then becomes,
to what extent is $\psi$ part of this new code subspace?
We will see that there is an associated area operator for this new code subspace, which in particular derives from  $(E_1 \otimes E_2)^\Phi$, and we will argue that the order
parameter is exactly the condition of quantum minimality - the two code subspaces define possible quantum extremal regions that are here causally related, and to pick the correct one we need to minimize the generalized entropy. 

\subsection{Split code subspaces}
\label{sec:splitcode}

We now introduce new code subspaces, that are derived from the previous subspace $V(\mathscr{H})$, and are based around the split state $S$ on the boundary Hilbert space.\footnote{There is 
a real sense in which one should think of this as corresponding to a new bulk geometry, that is constructed from data of the old bulk geometry. Similar in vein to the constructions of the canonical purification \cite{Dutta:2019gen} and in particular Engelhardt-Wall's geometry associated to non-minimal HRT surfaces \cite{Engelhardt:2018kcs}.} For boundary 2d QFTs on a circle
there are two such subspaces since the complement of two intervals is two intervals. In the next subsection we will focus on the 2d case, but
here we will concentrate on the the split code subspace associated to $\mathcal{M}_1 \otimes \mathcal{M}_2$. 
Define the subspace:
\begin{equation}
\mathscr{K} \supset V_S(\mathscr{H}_S) = \mathcal{U}^\dagger V \otimes V (\mathscr{H} \otimes \mathscr{H})
=  (\Theta')^\dagger  V (\mathscr{H})
\end{equation}
That is $V_S =  (\Theta')^\dagger V$ is the isometry that relates this new split bulk Hilbert space to the physical Hilbert space
and we have equality $\mathscr{H}_S = \mathscr{H}$, although we will continue to label this $\mathscr{H}_S$ to make it clear this is a different code subspace. 

 This subspaces contains the split state $S \in V_S(\mathscr{H}_S)$, and standardly reconstructable algebras $\mathcal{N}_1 \subset \mathcal{B}(\mathscr{H}_S)$ from
$\mathcal{M}_1$ and $\mathcal{N}_2 \subset \mathcal{B}(\mathscr{H}_S)$ from
$\mathcal{M}_2$. Where $\beta_1^S(\mathcal{N}_1) = \mathcal{N}_1^\beta$ and $\beta_2^S(\mathcal{N}_2) = \mathcal{N}_1^\beta$
and in particular $\beta_i^S = \beta_i$. These are not hard to construct using the existence of $\mathcal{N}_{1,2}$, and the fact that $\Theta' \in (\mathcal{N}_1^\beta \vee \mathcal{N}_2^\beta)'$.
Furthermore one can show that (see Eq~\ref{asplit} below):
\begin{equation}
\alpha_i^S(m_i) = V_S^\dagger m_i V_S = V^\dagger m_i V = \alpha_i(m_i)
\end{equation}
Thus the conditional expectations satisfy $E_i^S = E_i$ for $i=1,2$.
Note that we do not have $(\beta_i^S)' \neq (\beta_i)'$ and similarly $(\alpha_i^S)' \neq (\alpha_i)'$. 

Now the utility of the split state is the following simple set of result:

\begin{lemma}
\label{lemsplit2}
\begin{itemize}
\item[(i)] On the split code subspace the aglebra $\mathcal{N}_1 \vee \mathcal{N}_2 \subset \mathcal{B}(\mathscr{H}_S)$ is standardly c-reconstructable from $\mathcal{M}_1 \vee \mathcal{M}_2$.
\item[(ii)] Furthermore the conditional expectation $E_{12}^S$ satisfies:
\begin{equation}
E_{12}^S = (E_1 \otimes E_2)^\Phi
\end{equation}
and the area operator satisfies:
\begin{equation}
\mathcal{L}_{(\mathcal{N}_1 \vee \mathcal{N}_2)^{\beta^S}}
= \mathcal{L}_{\mathcal{N}_1^\beta} +  \mathcal{L}_{\mathcal{N}_2^\beta}
\end{equation}
where these area operators are those associated to the reconstructions $\mathcal{N}_{1,2}$.
\item[(iii)] States that are fixed on the original code subspace with $\rho = \rho \circ E_{12}$ have the property that:
\begin{equation}
S_{\mathcal{M}_1 \wedge (\mathcal{N}_1^\beta)'}( \rho) 
+ S_{\mathcal{M}_2 \wedge (\mathcal{N}_2^\beta)'}( \rho) = \rho\left(\mathcal{L}_{(\mathcal{N}_1 \vee \mathcal{N}_2)^{\beta^S}}\right)
\end{equation}
(even if these are not on the split code subspace $ \rho \neq \rho \circ E_{12}^S$. )  In the purely infinite setting we must assume here that the relative commutants are a discrete
sum of type-I factors.

\end{itemize}
\end{lemma}

\begin{proof} 
(i) We first need to construct $\beta_{12}^S$ satisfying the usual
properties. Consider the map:
\begin{equation}
\beta_{12}^S(n_1 n_2)
\equiv \Phi( \beta_{1}(n_1) \otimes \beta_{2}(n_2) ) \in \mathcal{M}_1 \vee \mathcal{M}_2 \,, \qquad n_i \in \mathcal{N}_i
\end{equation}
where:
\begin{align}
\beta_{12}^S(n_1 n_2) V_S \left| \psi \right>
&=  \mathcal{U}^\dagger \beta_{1}(n_1) \otimes \beta_{2}(n_2) 
\mathcal{U}( \Theta')^\dagger V \left| \psi \right> \\
&=\mathcal{U}^\dagger \beta_{1}(n_1) \otimes \beta_{2}(n_2) 
V \otimes V \mathcal{U}_{\mathscr{H}} \left| \psi \right> \\& =
\mathcal{U}^\dagger
V \otimes V  \mathcal{U}_{\mathscr{H}} n_1 n_2 \left| \psi \right>
\\& =
V_S n_1 n_2 \left| \psi \right>
 \,, \qquad  \left| \psi \right> \in \mathscr{H}
 \label{vb12}
\end{align}

This map extends linearly to sums and products of $n_i$. Thus by the split property, it extends to the full $ \mathcal{N}_1 \vee \mathcal{N}_2$
as an embedding and satisfying \eqref{vb12} for the full algebra.

Also, clearly for $m_i \in \mathcal{M}_i$:
\begin{align}
\label{asplit}
\alpha_{12}^S(  m_1 m_2 ) &= V_S^\dagger  \mathcal{U}^\dagger m_1\otimes m_2  \mathcal{U} V_S
 = \mathcal{U}_{\mathscr{H}}^\dagger V^\dagger m_1 V \otimes V^\dagger m_2 V \mathcal{U}_{\mathscr{H}}  
\\&= \Phi_{\mathscr{H}} ( \alpha_1(m_1) \otimes \alpha_2(m_2)) = \alpha_1(m_1) \alpha_2(m_2)
\end{align}
which is then clearly in $\mathcal{N}_1 \vee \mathcal{N}_2$. This extends to full algebra $\mathcal{M}_1 \vee \mathcal{M}_2$ via the split property.
It is faithful since $\alpha_i$ are faithful. 

Thus $\mathcal{N}_1 \vee \mathcal{N}_2 \subset \mathcal{B}(\mathscr{H}_S)$ is standardly c-reconstructable from $\mathcal{M}_1 \vee \mathcal{M}_2$ on the split code subspace
(with $V_S$). 

(ii) The conditional expectation can be computed as:
\begin{equation}
E_{12}^S (m_1 m_2) \equiv \beta_{12} \circ \alpha_{12} (m_1 m_2)= \Phi (E_1(m_1) \otimes E_2(m_2)) = \Phi \circ E_1 \otimes E_2 \circ \Phi^{-1}(m_1 m_2)
\end{equation}
which also extends as expected to $\mathcal{M}_1 \vee \mathcal{M}_2$. Note that the relative commutant in this case is:
\begin{equation}
(\mathcal{M}_1 \vee \mathcal{M}_2) \wedge (\mathcal{N}_1^\beta)' \wedge (\mathcal{N}_2^\beta)'
= (\mathcal{M}_1 \wedge (\mathcal{N}_1^\beta)' ) \vee ( \mathcal{M}_2) \wedge (\mathcal{N}_2^\beta)' )
\end{equation}
so that the state on the relative commutant factoraizes:
\begin{equation}
E_{12}^S(n_1^c n_2^c) = \left( \sum_a \pi_a^1 \chi^1_a (n_1^c) \right) \left( \sum_b \pi_b^2 \chi^2_b (n_2^c)\right) \,, \qquad n_i \in  \mathcal{M}_i \wedge (\mathcal{N}_i^\beta)' 
\end{equation}
where $\pi_a^1 \in Z(\mathcal{N}_1^\beta)$ and $\pi_b^2 \in Z(\mathcal{N}_2^\beta)$ are a complete set of minimal central projectors
and $\chi^1_a, \chi^2_b$ are the states on the relative commutant that uniquely determine $E_{12}^S$.
And since $E_1 = E_{12}|_{\mathcal{M}_1} = E_{12}^S|_{\mathcal{M}_1}$ and $E_2 = E_{12}|_{\mathcal{M}_2} = E_{12}^S|_{\mathcal{M}_2}
$ we see that the states $\chi^i$ defined above are the same as those associated to $E_i$ for the reconstruction of $\mathcal{N}_i$ by $\mathcal{M}_i$ on the non-split code. 
Since the entropies on the relative commutant must add the associated area operator adds in the obvious way.

(iii) If $\rho = \rho \circ E_{12}$ then it is also invariant under restriction $\rho = \rho \circ E_1$. The result follows since the area operator is additive. 
\end{proof}

This Lemma also allows us to compute explicitly the right hand side of \eqref{mid}, for a state $\rho = \rho\circ E_{12}$, and in the type-I setting.
We find again \eqref{eqn}

We then give the following characterization of additivity non-violations:
\begin{lemma} 
\label{lem3}
The following statements are equivalent:
\begin{itemize}
\item[(i)]  $V_S (\mathscr{H}_S ) \cap V(\mathscr{H}) \neq \emptyset $, and the intersection contains a cyclic and separating vector on either code sub-space for $\mathcal{N}_1 \vee \mathcal{N}_2$.
\item[(ii)] $\mathcal{N}_1 \vee \mathcal{N}_2 = \mathcal{N}_{12}$.
\item[(iii)] $\Theta' = 1$ or in other words the unitary implementation of the split property commutes with the projection to the code subspace:
\begin{equation}
 \mathcal{U} V  = V \otimes V \mathcal{U}_{\mathscr{H}} 
\end{equation}
and thus $V_S (\mathscr{H}_S ) = V(\mathscr{H})$.
\end{itemize}
\end{lemma}
\begin{proof}
(i) $\implies$ (ii). 
Let $\xi = V_S \zeta_S = V \zeta$ be such a cyclic and separating vector common to both code subspace. 
So:
\begin{equation}
\omega_{V \zeta}|_{\mathcal{N}_1^\beta \vee \mathcal{N}_2^\beta} = \omega_{V_S \zeta_S}|_{\mathcal{N}_1^\beta \vee \mathcal{N}_2^\beta}
= \omega_{V \zeta_S}|_{\mathcal{N}_1^\beta \vee \mathcal{N}_2^\beta}
\end{equation}
Or $(\omega_\zeta - \omega_{\zeta_S})|_{\mathcal{N}_1 \vee \mathcal{N}_2} = 0$. Thus there exists a unitary $u' \in (\mathcal{N}_1 \vee \mathcal{N}_2)'$
such that $\zeta_S = u' \zeta$. But then:
\begin{equation}
V_S u' \left| \zeta \right> = 
(\beta'_{12})^S(u') V_S \left| \zeta \right>  
= (\beta'_{12})^S(u') (\Theta')^\dagger V \left| \zeta \right>  =  V \left| \zeta \right>
\end{equation}
Now by the cyclic property for $\mathcal{N}_1$ it must be that $\mathcal{N}_1^\beta \vee \mathcal{N}_2^\beta$ generates
the full code subspace so
\begin{equation}
 V^\dagger \Theta' =  V^\dagger (\beta'_{12})^S(u')
\end{equation}
but since  $(\beta'_{12})^S(u') \in (\beta'_{12})^S( (\mathcal{N}_1 \vee \mathcal{N}_2)')\subset (\mathcal{M}_1 \vee \mathcal{M}_2)'$ we have that:
\begin{equation}
 V^\dagger \Theta'  ( m_{12}) (\Theta')^\dagger V = V^\dagger  m_{12} V\,, \qquad m_{12} \in \mathcal{M}_1 \vee \mathcal{M}_2
\end{equation}
and this implies that $\alpha_{12}^S = \alpha_{12}$. Thus $\alpha_{12}^S \circ \beta_{12}( \mathcal{N}_{12}) = \mathcal{N}_{12}$, 
but $\alpha_{12}^S$ maps to $\mathcal{N}_1 \vee \mathcal{N}_2$ so $\mathcal{N}_{12} \subset \mathcal{N}_1 \vee \mathcal{N}_2$.
We also have $\mathcal{N}_1 \vee \mathcal{N}_2 \subset \mathcal{N}_{12}$ (from Theorem~\ref{ewn}), so must have equality. 

(ii) $\implies$ (iii) This condition means the split state $\omega_S$ is in the code subspace since:
\begin{equation}
\omega_S \circ E_{12} (m_1 m_2) = \omega_{\Omega} \circ E_1(m_1)  \omega_{\Omega} \circ E_2(m_2)
=   \omega_{\Omega}(m_1)  \omega_{\Omega} (m_2) = \omega_S(m_1 m_2)
\end{equation}
where we used the form $E_{12} = (E_1 \otimes E_2)^\Phi$ appropriate to the case where $\mathcal{N}_{12}  
= \mathcal{N}_1 \vee \mathcal{N}_2$ (due to the split property the maps $\Phi_{\mathscr{H}}^{-1}\circ \alpha_{12} \circ \Phi$ and $\Phi^{-1} \circ \beta_{12} \circ \Phi_{\mathscr{H}}$ clearly extend to a form that preserves tensor product structure). Set $\omega_S \circ \beta_{12} = \omega_{\eta_S}$ for some vector $\eta_S \in \mathscr{H}$. 

 So we can pick $\left| S\right> = \phi' V \left| \eta_S\right>$ for $\phi' \in (\mathcal{M}_1 \vee \mathcal{M}_2)'$. We must have:
\begin{equation}
\Theta' \phi' V \left| \eta_S\right> = V\mathcal{U}_{\mathscr{H}} \left| \eta_\Omega \right> \otimes \left| \eta_\Omega \right>
\end{equation}
but equating the linear functionals
on $\mathcal{N}_1^\beta \vee \mathcal{N}_2^\beta$ this implies that $u' \left| \eta_S \right> = \mathcal{U}_{\mathscr{H}} \left| \eta_\Omega \right> \otimes \left| \eta_\Omega \right>$
for some unitary  $u' \in (\mathcal{N}_1 \vee \mathcal{N}_2)'$. (Since $\Theta'$ and $\phi'$ both commute through the linear functional.)

Now consider the modular conjugation operators for the vectors $\left| \eta_\Omega \right>$ and $\left| \Omega \right>$ and
the respective algebras $\mathcal{N}_1 \vee \mathcal{N}_2$ and  $\mathcal{M}_1 \vee \mathcal{M}_2$.
From \eqref{modflow} the modular conjugation operator satisfy: $V J_{\mathcal{N}_1 \vee \mathcal{N}_2}
= J_{\mathcal{M}_1 \vee \mathcal{M}_2} V$ such that  the natural cones must similarly map to each other under $V$. That
is, $V(\mathscr{P}^\natural_{\mathcal{N}_1 \vee \mathcal{N}_2}) \subset \mathscr{P}^\natural_{\mathcal{M}_1 \vee \mathcal{M}_2}$ via 
\begin{equation}
V J_{\mathcal{N}_1 \vee \mathcal{N}_2} n J_{\mathcal{N}_1 \vee \mathcal{N}_2}  n \left| \eta_\Omega \right>
= J_{\mathcal{M}_1 \vee \mathcal{M}_2} \beta_{12}(n) J_{\mathcal{M}_1 \vee \mathcal{M}_2}  \beta_{12}(n) \left| \Omega \right>
\end{equation}
where $n \in \mathcal{N}_1 \vee \mathcal{N}_2$. But then both vectors $\phi' V\left| \eta_S\right>$ and $V \mathcal{U}_{\mathscr{H}} \left| \eta_\Omega \right> \otimes \left| \eta_\Omega \right> 
= \beta_{12}'(u') V \left| \eta_S \right>$ are in the natural cone $\in \mathscr{P}^\natural_{\mathcal{M}_1 \vee \mathcal{M}_2} $. Thus $\beta_{12}'(u') \in (\mathcal{M}_1 \vee \mathcal{M}_2)'$ must be equal to $\phi'$ which finally implies that  $\Theta' = 1$. 

(iii) $\implies$ (i) is trivial. 
\end{proof}

\subsection{Dual additive codes}
\label{sec:dualadditive}

We now use the previous sections results to construct a code that can live in two possible distinct code subspaces, with
distinct bulk reconstructable algebras.
A minimization condition will determine which algebra is reconstructable. 

If our disjoint regions arise from a 2d QFT then a reasonable assumption for the boundary algebras of such a theory (that of strong additivity \cite{Kawahigashi:1999jz}) guarantees that 
$(\mathcal{M}_1 \vee \mathcal{M}_2)' = \mathcal{M}_3 \vee \mathcal{M}_4$ for algebras $\mathcal{M}_{3,4}$. We will assume this and furthermore that
 $\mathcal{M}_3 \subset \mathcal{M}_4'$ is also a standard split inclusion. 
 The corresponding bulk algebras $\mathcal{N}_3$ and $\mathcal{N}_4$ will satisfy:
\begin{equation}
\mathcal{N}_1 \vee \mathcal{N}_2 \subset \mathcal{N}_{12} \subset (\mathcal{N}_{34})' \subset (\mathcal{N}_3 \vee \mathcal{N}_4)'
\end{equation}
but we do not have equality. In fact if we have equality $\mathcal{N}_1 \vee \mathcal{N}_2 = (\mathcal{N}_3 \vee \mathcal{N}_4)'$ 
in this situation then we would not have addivity violations for either $12$ or $34$. Hence we will not consider such a situation.
Inspired by AdS/CFT we introduce the following constraint on such a four party code extending Definition~\ref{def:ncc}:
\begin{definition}[Dual additive codes]
 An n-party code is defined as:
 \begin{equation}
 \{ \mathcal{N}_1,\ldots,\mathcal{N}_n\,\mathcal{N}_1',\ldots,\mathcal{N}_n' \} \leftarrow
 \{ \mathcal{M}_1,\ldots,\mathcal{M}_n\,\mathcal{M}_1',\ldots,\mathcal{M}_n' \} \,: 
 \end{equation}
 where the order relation is such that $\{ \mathcal{M}_1, \ldots \mathcal{M}_n\}$ are all causally separated.\footnote{That is $\mathcal{M}_k \subset \mathcal{M}_1'\,\ldots \mathcal{M}_{k-1}',\mathcal{M}_{k+1}'
 ,\ldots \mathcal{M}_n'$ for all $k=1, \ldots n$. }
 A 4-party code with $ \mathcal{M}_1 \vee \mathcal{M}_2 = \mathcal{M}_3' \wedge \mathcal{M}_4'$ 
 is called dual-additive if  there is are algebras $\mathcal{N}_{12}$ that is standardly c-reconstructable from $\mathcal{M}_1 \vee \mathcal{M}_2$ and $\mathcal{N}_{34}$ 
 that is standardly c-reconstructable from $\mathcal{M}_3 \vee \mathcal{M}_4$. Where we assume both $\mathcal{M}_1 \vee \mathcal{M}_2$ and $\mathcal{M}_3 \vee \mathcal{M}_4$ are standardly split and where additionally one (and only one) of the following is satisfied: $\mathcal{N}_1 \vee \mathcal{N}_2 = \mathcal{N}_{12}$ 
{\bf or} $\mathcal{N}_3 \vee \mathcal{N}_4 = \mathcal{N}_{34}$.\footnote{Note that $\mathcal{N}_{12} = \mathcal{N}_{34}'$ by the definition of reconstructable.}
We call a net of error correcting codes dual additive if any 4-party sub-code, with $\mathcal{M}_1 \vee \mathcal{M}_2 = \mathcal{M}_3' \wedge \mathcal{M}_4'$, is dual additive.
\end{definition}

\begin{theorem}
\label{pt}
In  the type-I setting, a net of error correcting codes is dual additive, iff it satisfies the following quantum minimality condition on any 4-party sub-code with $\mathcal{M}_1 \vee \mathcal{M}_2 = \mathcal{M}_3' \wedge \mathcal{M}_4'$:
\begin{equation}
\label{t1}
S_{\mathcal{M}_{1} \vee \mathcal{M}_2}(\rho \circ \alpha) = \min_{\mathcal{N}\in\{\mathcal{N}_{1}\vee \mathcal{N}_2,\,\mathcal{N}_{3}' \wedge\, \mathcal{N}_4' \}}  \left( \rho(\widehat{A}_{\mathcal{N}})+ S_\mathcal{N}(\rho) \right)
\end{equation}
for all $\rho \in \mathcal{B}(\mathscr{H})_\star$. Furthermore the bulk sub-algebra that achieves the minimum is standardly c-reconstructable from $\mathcal{M}_1 \vee \mathcal{M}_2$.
\end{theorem}

\begin{proof}
Pick some purification $\rho= \omega_{\psi}|_{\mathcal{N}_3' \wedge \mathcal{N}_4'}$. 
The entropies can be computed using \eqref{eqn} and purity of $\psi$
such that \eqref{t1} becomes:
\begin{equation}
\label{pinf2}
0 = \min\{ S_{\rm rel}(\omega_\psi \circ \alpha_{12} | \omega_\psi \circ \alpha_{12} \circ  (E_1 \otimes E_2)^{\Phi_{12}}),\, S_{\rm rel}(\omega_\psi \circ \alpha_{34} | \omega_\psi \circ \alpha_{34} \circ (E_3 \otimes E_4)^{\Phi_{34}} )\}
\end{equation}
If the 4-party code is dual additive, we know by Lemma~\ref{lem3} that either $V(\psi) \in V_{S_{12}} (\mathscr{H}_{S_{12}})$
or $V(\psi) \in V_{S_{34}} (\mathscr{H}_{S_{43}})$ where $12$ or $34$ label the two different split codes. 
Thus either $\omega_{\psi} \circ \alpha_{12} \circ (E_1 \otimes E_2)^{\Phi_{12}} = \omega_{\psi} \circ \alpha_{12}$
or  $\omega_{\psi} \circ \alpha_{34} \circ (E_3 \otimes E_4)^{\Phi_{34}} = \omega_{\psi} \circ \alpha_{34}$.
 and so the result follows. 

For the converse statement we pick some $\rho = \omega_\psi$ for a jointly cyclic and separating vector $\psi$ such that the vanishing of relative entropy in \eqref{pinf2} tells us that either 
 $V(\psi) \in V_{S_{12}} (\mathscr{H}_{S_{12}})$
or $V(\psi) \in V_{S_{34}} (\mathscr{H}_{S_{43}})$ and by Lemma~\ref{lem3} this implies that either $\mathcal{N}_{12} = \mathcal{N}_1 \vee  \mathcal{N}_2$ or 
$\mathcal{N}_{34} = \mathcal{N}_3 \vee  \mathcal{N}_4$. 

The reconstruction statements follow immediately. \end{proof}

Note that it is possible to treat  the purely infinite case and the necessary and sufficient condition is simply
\eqref{pinf2}. 

We conjecture that the dual additive code describes two dimensional holographic CFTs. Indeed the HRT formula with quantum corrections satisfies \eqref{t1} as has been proven using
replica methods for a code subspace based around global AdS with small backreaction. So Theorem~\ref{pt} essentially demonstrates this. 

Furthermore we conjecture that theories with a large central charge $c$, and a sparse spectrum
of operator dimensions furnish a dual additive net of complementary codes. In particular the code Hilbert space will involve a projection to the low lying primary operators, and indeed
in a 2d CFT such projections give rise to subnets. This conjecture is sufficiently vague, that future work will have to fill in the details of exactly what conditions to impose on the spectrum
of operators, how to take the limit $c \rightarrow \infty$ (if at all) and how many low lying operators to include in the code.
We mention a strong motivation for this conjecture comes from more traditional approaches to entanglement in QFT - that is the replica trick. 
The Renyi entropies of two disjoint intervals demonstrate a similar phase transition to entanglement entropy, and the case $n=2$ is provably controlled by the large $c$ and a sparse spectrum
assumptions \cite{Hartman:2014oaa}.

\subsection{An entropy bound}
\label{sec:hbound}

The resulting phase transitions governed by dual additive codes described by Theorem~\ref{pt} can naturally be thought of as arising from changing the boundary algebras. Note that under such a change, the two split codes change and so do the area operators. In fact it is reasonable that for very far separated algebras $\mathcal{M}_1$ and $\mathcal{M}_2$  associated to intervals on a spatial slice, the code $V$ 
must be the same as the split code $V_{S_{12}}$. For symmetry reasons, say in a 2d CFT, there must then be a transition to the other split code at $x=1/2$ where $x$ is the cross ratio of the end points of each interval. 

This outcome is however slightly awkward compared to the expectations of AdS/CFT.
The reason for this, as usual, stems from the fact that we are working with exact recovery.
This means that the two different split code subspaces cannot overlap and so we cannot ``continuously'' extrapolate from one reconstructable algebra to another by changing the state while remaining in some fixed code subspace. See for example \cite{Akers:2019wxj} where it was shown that exactly the transition we are trying to model can be achieved by changing the state.
In reality we expect a small non-perturbative overlap between the codes and these small exponentials can build up to such a transition. In particular the bulk entropy term in \eqref{t1} can become large and force such a transition, by overwhelming any area difference. 

In some sense the codes here are constrained by a bound more reminiscent of the QFT Bekenstein entropy bound as formulated in \cite{Casini:2008cr}.
This bound, which is derivable within QFT without gravity, simply limits the entropy of QFT states to not be large enough to violate various well motivated inequalities from semi-classical gravity. 
In this way it is simply not possible to build up large entropies that overwhelm the area terms. 
Along these lines we now discuss a holographic entropy bound that arises when working in a fixed code subspace 
and puts an important constraint on the existence of a dual additive code. 

In the type-$I$ setting, the dual additive condition implies for the sector with $\mathcal{N}_{34} = \mathcal{N}_3 \vee \mathcal{N}_4$ that
all entropies in the code subspace must satisfy:
\begin{equation}
\label{bb}
S_{\mathcal{N}_3' \wedge \mathcal{N}_4'}(\rho) \leq S_{\mathcal{N}_1 \vee \mathcal{N}_2}(\rho) 
+ \rho( \mathcal{A}_{\mathcal{N}_1} + \mathcal{A}_{\mathcal{N}_2} -\mathcal{A}_{\mathcal{N}_3}- \mathcal{A}_{\mathcal{N}_4})
\end{equation}
for all $\rho \in \mathcal{B}(\mathscr{H})$.
To draw out the consequence of this inequality consider states invariant under a bulk conditional expectation: $\rho = \rho \circ E_0$ where $E_0: \mathcal{N}_3' \wedge \mathcal{N}_4' \rightarrow \mathcal{N}_1 \vee \mathcal{N}_2$ for the inclusion:
\begin{equation}
\label{callback}
\mathcal{N}_1 \vee \mathcal{N}_2 \subset  \mathcal{N}_3' \wedge \mathcal{N}_4' 
\end{equation}
 The central state on the bulk relative commutant is determined by numbers $\lambda_{\gamma \alpha}$ (see for example \cite{Giorgetti:2018nji}):
\begin{equation}
E_0(\varpi^{34}_{\gamma}) \varpi^{12}_\alpha = \lambda_{\gamma\alpha}  \varpi^{12}_\alpha
\end{equation}
where $\varpi_\alpha^{12}$ and $\varpi_\gamma^{34}$ are central projectors
in $Z(\mathcal{N}_1) \cup Z(\mathcal{N}_2)$ and $Z(\mathcal{N}_3) \cup Z(\mathcal{N}_4)$
respectively. 
For an optimal bound we pick $E_0$ such that the corresponding states on the bulk relative commutant
for fixed $\varpi_\alpha^{12}$ and $\varpi_\gamma^{34}$ are maximally mixed
acting on an (assumed) finite dimensional Hilbert space with size $N_{ \gamma\alpha}$. Note that $\lambda_{\alpha \gamma} \geq 0$ and $\sum_\gamma \lambda_{\gamma \alpha} = 1$.
The state on the bulk relative commutant $ \mathcal{N}_1' \wedge \mathcal{N}_2' \wedge \mathcal{N}_3' \wedge \mathcal{N}_4'$ is then:
\begin{equation}
\rho(\cdot)= \rho\circ E_0(\cdot) =\bigoplus_{\gamma\alpha} \frac{\lambda_{\gamma\alpha} p_\alpha}{N_{\gamma\alpha}} {\Tr}_{\mathcal{H}_{\gamma\alpha}}(\cdot)
\end{equation}
where $p_\alpha = \rho(\varpi^{12}_\alpha)$. One can then compute \eqref{bb} explicity:
\begin{equation}
\sum_{\gamma \alpha} \lambda_{\gamma \alpha} p_\alpha (\ln N_{\gamma \alpha} - \ln ( \lambda_{\gamma\alpha}))
\leq \sum_{ \alpha} p_\alpha S(\chi_\alpha^{12}) - \sum_\gamma p_\gamma S(\chi_\gamma^{34})  
\end{equation}
and where we have put the  area operators/states for $12$ and $34$ together  $S(\chi_{\alpha = ab}^{12}) = S(\chi_{a}^{1}) + S(\chi_{b}^{2})$ etc. Maximize over $\lambda_{\gamma \alpha}$ and minimizing over $p_\gamma$
with the constraint that $\sum_\gamma \lambda_{\gamma \alpha}= 1$ and $\sum_\gamma p_\gamma=1$. We must have $\lambda_{\gamma\alpha} = 1/ n_{34}$ and $p_\gamma = \delta_{\gamma, \gamma_m}$
where $S(\chi_{\gamma_m}^{34})$ is the maximum such entropy and $n_{34}$ is the size of
the $34$ center.
Setting $p_\alpha = \delta_{\alpha,\alpha'}$ we get:
\begin{equation}
\label{cons}
\sum_{\gamma } \frac{1}{n_{34}}  \ln (N_{\gamma \alpha}n_{34})
\leq S(\chi_\alpha^{12}) - \max_{\gamma}( S(\chi_\gamma^{34})  )
\end{equation}
The left hand side is roughly the log of the size of the relative commutant for a fixed $\alpha$ sector.
This is a strong bound on the central spectrum of the area operator. It is also a strong bound on the size of the bulk relative commutant - and hence on the size
of the code subspace. In particular one notes that the entropies $S(\chi_\alpha^{12}) $ must all be larger than all the entropies $S(\chi_\gamma^{34})$.

For two intervals in a 2d CFT (and forgetting temporarily the fact that the type-I setting does not apply here), 
by symmetry we expect at $x=1/2$ the various entropies $S(\chi)$ will share a symmetry between $12 \leftrightarrow 34$ which then would imply by \eqref{cons} that
all the entropies in the central decomposition must be equal and the size of the relative commutant vanishes.
This is likely not a correct interpretation. In particular one might expect the size of the relative commutant to be roughly fixed
near the phase transition, and certainly not zero. 
Rather we expect our error correcting code to start to break down near $x=1/2$, showing strong deviations
from exact recovery and dual additivity. 

\section{Tensions with holography}

\label{sec:tensions}

We now discuss some issues interpreting a net of complementary codes as a model of holography. The conclusion will be that the exact recovery condition is too strong. We hope however that some of this rather nice and natural structure survives approximation. 

Our first result pertains to additivity of overlapping regions, and was first discussed in AdS/CFT in \cite{Kelly:2016edc}:
\begin{theorem}
\label{thmadd}
Assume that $\mathcal{N}_{1,2}$ are respectively standardly c-reconstructable from $\mathcal{M}_{1,2}$.
If there is a vector $V\eta$ which is cyclic for  $\mathcal{M}_{1} \wedge \mathcal{M}_2$
then $\mathcal{N}_1 \vee \mathcal{N}_2$ is c-reconstructable from $\mathcal{M}_1 \vee \mathcal{M}_2$. 
\end{theorem}
\begin{proof}
We know that $\mathcal{N}_1 \vee \mathcal{N}_2$ is reconstructable from $\mathcal{M}_1 \vee \mathcal{M}_2$  by the following argument.
By entanglement wedge nesting (Theorem~\ref{ewn}) since $\mathcal{M}_{1,2} \subset \mathcal{M}_{1} \vee \mathcal{M}_2$
it follows that $\mathcal{N}_1$ and $\mathcal{N}_2$ are reconstructable from $\mathcal{M}_{1} \vee \mathcal{M}_2$
so that $\alpha|_{\mathcal{M}_{1}' \wedge \mathcal{M}_2'} \in \mathcal{N}_1' \wedge \mathcal{N}_2'$. 

Now consider some operator $x \in \mathcal{N}_1' \wedge \mathcal{N}_2'$. We know that $\beta_1'(x) \in \mathcal{M}_1'$ 
and $\beta_2'(x) \in \mathcal{M}_2'$. But:
\begin{equation}
\label{preclude}
(\beta_1'(x)  - \beta_2'(x)) V\left| \eta \right> = V(x -x)\left| \eta \right> =0
\end{equation} 
Certainly $(\beta_1'(x)  - \beta_2'(x)) \in \mathcal{M}_1' \vee \mathcal{M}_2'$. So by the separating property of $V \eta$
for $ \mathcal{M}_1' \vee \mathcal{M}_2'$ it must be that $\beta_1'(x)  = \beta_2'(x) \in \mathcal{M}_2'$. But then:
\begin{equation}
\beta_{1}'(x) \in \mathcal{M}_1' \wedge \mathcal{M}_2'
\end{equation}
and so $\beta_1': \mathcal{N}_1' \wedge \mathcal{N}_2' \rightarrow  \mathcal{M}_1' \wedge \mathcal{M}_2'$ can be used in Theorem~\ref{thm1}
to show that $\mathcal{N}_1' \wedge \mathcal{N}_2'$ is reconstructable from $\mathcal{M}_1' \wedge \mathcal{M}_2'$. 
\end{proof}

\begin{corollary}
Assume that $\mathcal{N}_{1,2}$ are respectively standardly c-reconstructable from $\mathcal{M}_{1,2}$.
If there is a vector $V\eta$ which is separating for  $\mathcal{M}_{1} \vee \mathcal{M}_2$
then $\mathcal{N}_1 \wedge \mathcal{N}_2$ is c-reconstructable from $\mathcal{M}_1 \wedge \mathcal{M}_2$. 
\end{corollary}

\begin{figure}[h!]
\centering
\includegraphics[scale=.7]{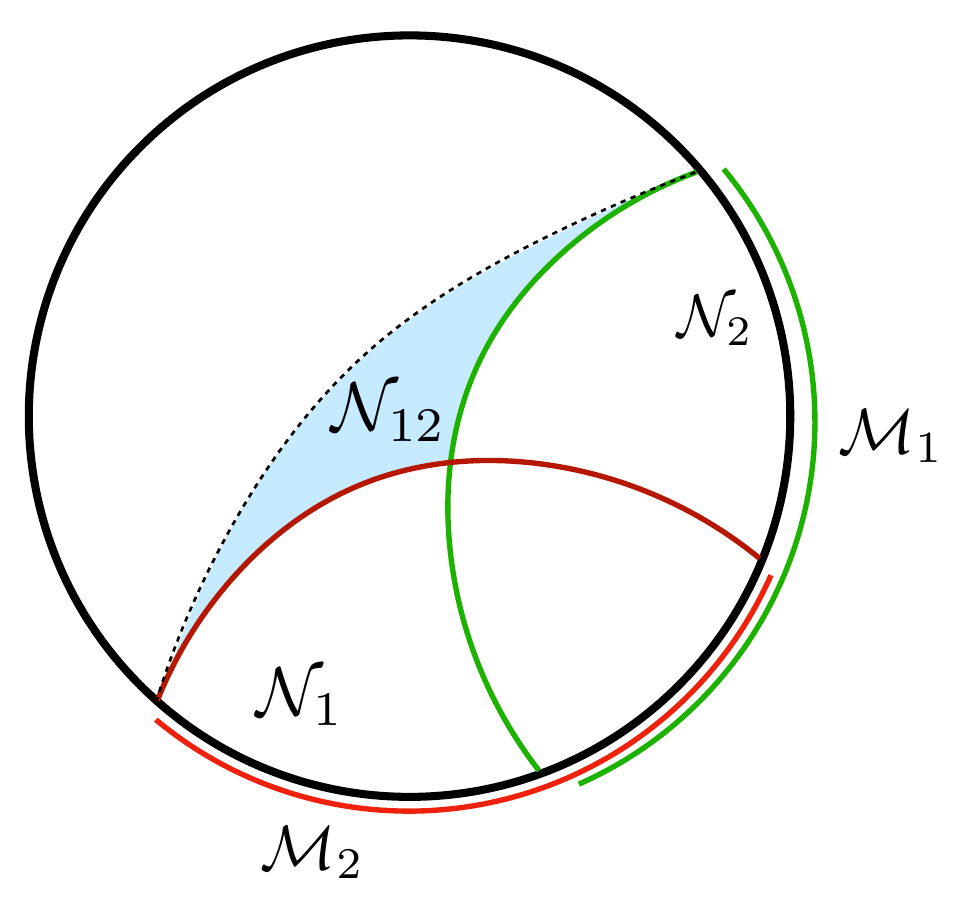}
\caption{\label{fig}  A cartoon of $AdS_3/CFT_2$ where we show a timeslice of the dual geometry describing the vacuum.
The regions are all timelike subregions of some fixed time slice, and the labelled algebras are associated to the causal completion of these regions
using respectively the bulk and boundary causal structure. Two intersecting boundary regions are not additive in the bulk due to the blue region. 
That is, in AdS/CFT, the entanglement wedge algebra for $\mathcal{M}_1 \vee \mathcal{M}_2$ is $\mathcal{N}_{12}$ which includes the blue region.
However this contradicts the assumptions that go into Theorem~\ref{thmadd} where the c-reconstructable algebra (assumed to be the entanglement wedge) is $\mathcal{N}_1 \vee \mathcal{N}_2$.}
\end{figure}

This is certainly not true in holography as can be seen from Figure~\ref{fig}.
In fact it was exactly this setting that motivated the error correction approach to holography in the first place \cite{Almheiri:2014lwa}.
The usual idea for how this works in holography is  that there are necessarily multiple reconstructions for the same operator.
However the cyclic and separating properties are so strong that one can preclude this possibility as in \eqref{preclude}.

One might give up the on the cyclic and separating property to save this model, however this seems like a very mild property and is certainly true
if the vacuum state is in the code subspace. We conclude that error correction in holography is necessarily approximate, and as discussed in \cite{Kelly:2016edc,Hayden:2018khn} using this
example, the approximate nature is important even in the absence of black hole horizons. 
Indeed the cyclic and separating statement does not seem very stable to approximation (at least in certain contexts.)
So for example if we can only approximately reconstruct the operator as $x_1,x_2$ respectively in
\eqref{preclude} then $V(x_1 - x_2) \left| \eta \right> \neq 0$ should be small but non-zero. The conclusion of this theorem 
no longer hold, and it is not clear that to what extent it might approximately hold. 

It seems reasonable that in certain situations this constraint on overlapping boundary regions, will also heavily constrain phase transitions between
two disjoint regions of the kind studied in Section~\ref{disjoint}. We have not managed to make a precise statement. However we mention here that there are some reasonable extra conditions that a general class of conformal field theory subnets $N \rightarrow M$ satisfies, that do spell trouble for these phase transitions \cite{Kawahigashi:1999jz}.
These assumptions are not guaranteed to be important for holography.  Perhaps the most troubling is strong additivity for $N$ (we already effectively assumed it for $M$) which is a much stronger version of Theorem~\ref{thmadd} where the (closure of the) regions are just touching.
Indeed strong additivity implies that the inclusion \eqref{callback} is irreducible/singular (so the relative commutant is trivial) and all the bulk algebras are factors. Assuming the inclusion \eqref{callback} has finite index then implies the index is constant over the code. Trivial bulk relative commutant is bad enough - the region between two extremal RT surfaces should be associated to a non-trivial algebra. Constant index seems to preclude any phase transitions in the entropy. 
 
\section{Conclusions}

There are many open questions and generalizations to investigate and we look forward to attempting some of these.

\acknowledgments

We thank Chris Akers, Fikret Ceyhan, Netta Engelhardt, Min Li and Pratik Rath for useful discussions. 
This work is partially supported by the Air Force Office of Scientific Research under award number FA9550-19-1-0360
and by the National Science Foundation under Grant No. NSF PHY- 1748958.

\appendix

\section{Proof of Theorem~\ref{thm1}}
\label{app:proof}

\begin{proof} The cyclic and separating case, for (ii) $\leftrightarrow$ (iii) was studied in \cite{accardi1982conditional}.  We will follow this paper very closely, but allow for the more
general assumptions in the theorem. 

(i) $\implies$ (ii) Consider some unitary $u \in \mathcal{N}$. Set $\psi = u \phi$, then the linear functionals on $\mathcal{N}'$
induced by $V\psi$ and $V\phi$ coincide implying that:
\begin{equation}
\label{phiX}
\left< \phi \right|  X \left| \phi \right> =0  \,, \qquad X \equiv  u^\dagger V^\dagger m' V u - V^\dagger m' V  
\in  \mathcal{B}(\mathscr{H})
\end{equation}
Any element of the predual $\varrho \in \mathcal{B}(\mathscr{H})_\star$
is a linear combination of four positive elements $\varrho_{k} \in (\mathcal{B}(\mathscr{H})_\star)^+$
for $k=1, \ldots,4$. Each of these elements can be represented
\cite{bratteli2012operator} as a (potentially) infinite sequence of vectors $\xi_i$ with $\sum_i \left< \xi_i \right| \left. \xi_i \right> < \infty$
and $\varrho_k(\cdot) = \sum_i \left< \xi_i \right| \cdot  \left| \xi_i \right>$.
Using \eqref{phiX} we find $\varrho_k(X) = 0$ and thus:
\begin{equation}
\varrho  \left( X  \right) = 0
\end{equation}
This implies that $X=0$ since $\mathcal{B}(\mathscr{H})$ has a unique correspondence with the linear functionals
on $ \mathcal{B}(\mathscr{H})_\star$ (the dual of the pre-dual.) 
Thus:
\begin{equation}
u^\dagger V^\dagger m' V u - V^\dagger m' V = 0 \quad \implies \quad \left[ V^\dagger m' V , u \right] = 0
\end{equation}
Thus $\alpha'(\mathcal{M}') = V^\dagger \mathcal{M}' V \subset \mathcal{N}'$. $\alpha'$ is obviously normal unital and completely positive.

For the bracketed statement [$ \ldots $] we consider $\omega_{\eta} \circ \alpha'$ which is faithful by assumption. So if $\alpha'(m' ) =0$ for $m' \geq 0$ then 
$\omega_{\eta} \circ \alpha'(m') = 0$ which implies $m'=0$.  

(ii) $\implies$ (i) The two purifications must be related via $\psi= u \phi$ for some partial isometry, with appropriate support, $u \in \mathcal{N}$.  
A short computation shows:
\begin{equation}
\left< \phi \right| u^\dagger V^\dagger m' V u \left| \phi \right>
= \left< \phi \right| u^\dagger V^\dagger m' \beta(u) V \left| \phi \right>
=  \left< \phi \right| u^\dagger u V^\dagger m' V \left| \phi \right> = \left< \phi \right| V^\dagger m' V \left| \phi \right>
\end{equation}
or $\left. \omega_{V\psi} \right|_{\mathcal{M}'}  = \left. \omega_{V\phi} \right|_{\mathcal{M}'}$ as required.

(iii) $\implies$ (ii) The injective *-homomorphism property implies that  $\mathcal{N}^\beta \equiv \beta(\mathcal{N})$ is a von Neumann algebra acting
on the Hilbert space $\beta(1) \mathscr{K}$ where $\beta(1)\in \mathcal{M}$ is a projection. We have that $\mathcal{N}^\beta \subset \beta(1) \mathcal{M} \beta(1) \subset \mathcal{M}$. 
We can derive from \eqref{chan} that $\beta(1) VV^\dagger = VV^\dagger$ and $[\beta(n), VV^\dagger ] = 0$ for all $n$. 
Thus that the code subspace satisfies $V(\mathscr{H}) \subset \beta(1) \mathscr{K}$ such that 
the projector $e \equiv VV^\dagger$ acts within $\beta(1) \mathscr{K}$ and satisfies $e \in (\mathcal{N}_\beta)'$ (the commutant algebra on $\beta(1) \mathscr{K}$). 

We see that for all $\nu' \in (\mathcal{N}_\beta)'$ and $n \in\mathcal{N}$:
\begin{equation}
\left[ V^\dagger \nu' V , n \right] = \left[ V^\dagger \nu' V , V^\dagger\beta (n) V \right]
= V^\dagger ( \nu' e \beta (n)  - \beta(n)  e  \nu' ) V = 0
\end{equation}
so that $V^\dagger \mathcal{M}' V  = V^\dagger \mathcal{M}' \beta(1) V \subset V^\dagger (\mathcal{N}_\beta)' V  \subset \mathcal{N}'$. Also $\alpha'$ obviously has all the other properties of a unital quantum channel.

The bracketed statement [ $\ldots$] follows the same argument as in (i) $\implies$ (ii). 

(ii) $\implies$ (iii). Let us consider the case where there exists a vector $\eta \in \mathscr{H}$ such that $ \mathcal{M}' V \left| \eta \right>$ is dense in the
code subspace $V V^\dagger$.  Or in other words:
\begin{equation}
\label{stas}
VV^\dagger \leq \pi_{\mathcal{M}}(V \eta)
\end{equation}
Note that $\overline{V^\dagger \mathcal{M}'V \eta} \subset \overline{\mathcal{N}' \eta}$ so that:
\begin{equation}
1 = V^\dagger \pi_{\mathcal{M}}(V \eta) V \leq  \pi_{\mathcal{N}}( \eta) \leq 1
\end{equation}
so this assumption \eqref{stas} entails $  \pi_{\mathcal{N}}( \eta) = 1$. To ease the notation set $\pi_{\mathcal{M}'}(V \eta) = \pi'$ and $\pi_{\mathcal{M}}(V \eta) =\pi$ for the rest of the proof.

Then we can construct the dual map $\beta(n)$ as follows. Consider an arbitrary positive element $n_+ \in \mathcal{N}_+$. Define the unnormalized normal state $\rho \in \mathcal{M}'_\star$:
\begin{equation}
\rho(m') \equiv \left( n_+ \left| \eta \right>, \alpha'(m') \left| \eta \right> \right) = \left( n_+ \left| \eta \right>, V^\dagger m' V\left| \eta \right> \right) 
\end{equation}
This state is dominated by $\omega_{V\eta}(\cdot) \in \mathcal{M}'_\star$ since:
\begin{equation}
\rho(m_+')  =  \left( \alpha'(m'_+)^{1/2} \left| \eta \right>, n_+ \alpha'(m'_+)^{1/2} \left| \eta \right> \right) \leq \| n_+ \| \omega_{V\eta}(m_+') 
\end{equation}
So we can apply the commutant Radon-Nikodym theorem (see for example \cite{stratila2019lectures}) to conclude that there exists a positive element 
$\beta(n_+) \in \mathcal{M}$ (possibly non-unique) such that:
\begin{equation}
\rho(m') = \left( \beta(n_+) V \left| \eta \right>, m' V \left| \eta \right> \right)
\end{equation}
Note that any $\beta(n_+)$ that satisfies this equation can be replaced by $\beta(n_+)  \rightarrow \pi \beta(n_+) \pi $ and it still satisfies this equation. 
So in this way we pick $\beta(n_+) \in \pi \mathcal{M} \pi \subset \mathcal{M}$. Here $\pi \mathcal{M} \pi$ is a von Neumann algebra when taken to act on $\pi \mathscr{K}$. 

Extending this away from the positive part by linearity we find a positive map between von Neumann algebras $\beta(n)$ that satisfies:
\begin{equation}
\label{defbeta}
 \left( V n \left| \eta \right>, m' V\left| \eta \right> \right)  = 
 \left( \beta(n) V \left| \eta \right>, m' V \left| \eta \right> \right)
\end{equation}

Using the density of $m' V \left| \eta \right>$ we have:
\begin{equation}
 V n \left| \eta \right>= \pi V n \left| \eta \right>  = \pi  \beta(n) V \left| \eta \right>
\end{equation}
where we have used \eqref{stas}.
Act with $m'$ followed by $VV^\dagger$ we have:
\begin{equation}
V  n V^\dagger m' V \left| \eta \right> = V V^\dagger  \beta(n) m' V \left| \eta \right>
\end{equation}
and again by the density $m' V \left| \eta \right>$:
\begin{equation}
\label{vvv}
V  n V^\dagger   = V V^\dagger  \beta(n)
\end{equation}
where we used the fact that $\beta(n) \in \pi \mathcal{M} \pi$.  Taking daggers and acting with $V$ from the left we derive the required \eqref{chan}. 
By taking the dagger of \eqref{vvv} we can show
that $\left[ V V^\dagger, \beta(n) \right]=0$. This is a *-homomorphism since:
\begin{equation}
\beta(n_1 n_2) V=  V n_1 n_2
=  Vn_1V^\dagger V n_2 = \beta(n_1)V V^\dagger \beta(n_2) V
= \beta(n_1) \beta(n_2) V
\end{equation}
Applying this equation to $\left| \eta \right>$ and acting from the left with $m'$:
\begin{equation}
\beta(n_1 n_2) m' V\left|\eta \right> 
= \beta(n_1) \beta(n_2) m' V\left|\eta \right>
\end{equation}
and using the density $\overline{ m' V \left| \eta \right>} \rightarrow \pi$ so that:
\begin{equation}
\label{hom}
\beta(n_1 n_2) \pi = \beta(n_1)\beta(n_2) \pi\quad \implies  \beta(n_1 n_2)  = \beta(n_1)\beta(n_2)
\end{equation} 
where $\pi$ can be removed since $\beta(n) \in \pi \mathcal{M} \pi$. 
A similar argument establishes that $\beta(1) = \pi$. The dual map $\beta$ is faithful since:
\begin{equation}
0 = \beta(n_+) V = V n_+
\end{equation}
implies that $n_+ =0$. Thus $\beta$ is injective.  
Normality of the map $\beta(n)$ is argued for as follows (as in \cite{accardi1982conditional}). Take a norm bounded increasing net of positive operator $n_\alpha$
such that $\sup_{\alpha} n_{\alpha} = n_0$. Then consider $\beta(n_\alpha)$ which are also norm bounded and increasing by positivity of $\beta$. These must
converge (say weakly) to a positive element $\sup_{\alpha} \beta(n_{\alpha})$ in $\pi \mathcal{M} \pi$ that we label as $\beta_0$. Then by \eqref{defbeta}:
\begin{align}
\left( \beta(n_0) V \left| \eta \right>, m' V\left| \eta \right> \right) &= \left( V n_0 \left| \eta \right>, m' V\left| \eta \right> \right)
= \sup_\alpha \left( V n_\alpha \left| \eta \right>, m' V\left| \eta \right> \right) \\
&= \sup_\alpha \left(\beta(n_\alpha ) V \left| \eta \right>, m' V\left| \eta \right> \right)
= \left( \beta_0 V \left| \eta \right>, m' V \left| \eta \right> \right)
\end{align}
for all positive $m' \in \mathcal{M}'$. Using a similar density argument as for \eqref{hom} we have $\beta(n_0)  = \beta_0$ which establishes normality of this map. 
Finally complete positivity 
follows since for all $k \in \mathbb{Z}_{>0}$ elements $m_i' \in \mathcal{M}'$ and $n_i \in \mathcal{N}$
\begin{equation}
\sum_{ij} \left( m_i' V \left| \eta \right>, \beta( n_i^\dagger n_j) m_j' V \left| \eta \right> \right)
= \sum_{ij} \left( n_i V \left| \eta \right>,  V^\dagger (m_i')^\dagger m_j' V n_j  \left| \eta \right> \right)
\end{equation}
The right hand side is clearly positive and the left hand side can be used to approximate:
\begin{equation}
\left< \Phi \right| \sum_{ij} \beta( n_i^\dagger n_j) \otimes \left| i \right> \left< j \right| \left| \Phi \right>
\end{equation}
for any $\Phi \in \pi \mathscr{H} \otimes \mathscr{H}_k$ where $\mathscr{H}_k$ is a $k$-dimensional Hilbert space. 
Thus $\beta : \mathcal{N} \rightarrow \pi \mathcal{M} \pi$ is completely positive which implies complete positivity when extended to $\mathcal{M}$. 

We can now extend this proof to the more general case  where we do not assume the existence of $\eta$ satisfying  \eqref{stas}. 
Enlarge the Hilbert spaces by tensoring in a reference. Now $V \otimes 1_R$ acts between $\mathscr{H} \otimes \mathscr{H}_R
\rightarrow \mathscr{K} \otimes \mathscr{H}_R$ and
we extend the quantum channel $\alpha'$ to act on operators $\alpha' : M' \rightarrow  N'$ with the new algebras:
\begin{align}
N' = \mathcal{N}' \otimes \mathcal{B}(\mathscr{H}_R) \subset    \mathcal{B}(\mathscr{H}) \otimes \mathcal{B}(\mathscr{H}_R) \qquad (N = \mathcal{N} \otimes 1_R ) \\
M' = \mathcal{M}' \otimes \mathcal{B}(\mathscr{H}_R) \subset   \mathcal{B}(\mathscr{K}) \otimes \mathcal{B}(\mathscr{H}_R)  \qquad (M = \mathcal{M} \otimes 1_R )
\end{align}
and we define $\alpha'( m' \otimes e_{ij}) = \alpha'(m') \otimes e_{ij}$ where $i$ is a basis for $\mathscr{H}_R$. 
We will take $\mathscr{H}_R$ to be isomorphic $\mathscr{H}$.
Now we pick a vector $\left| \Phi \right> \in \mathscr{H} \otimes \mathscr{H}_R$ which is cyclic and separating for $1_{\mathscr{H}} \otimes \mathcal{B}( \mathscr{H}_R)$,
this is always possible (assuming the $\sigma$-finite condition.)
It is clear that $\pi_{M}( (V \otimes 1_R) \Phi) \geq \pi_{\mathcal{B}(\mathscr{K}) \otimes 1_R}( (V \otimes 1_R)\Phi) = V^\dagger V \otimes 1_R$. 
So if we consider this later space the ``code subspace'' we can run exactly the same argument as above where \eqref{stas} is satisfied for $\Phi$ and $M$. The dual channel is:
\begin{equation}
B : \mathcal{N} \otimes 1_R \rightarrow  \mathcal{M} \otimes 1_R
\end{equation}
from which we can extract $\beta(n) \otimes 1_R$. $B$ and hence $\beta$ satisfies the usual properties of a normal injective *-homomorphism and the required equation \eqref{chan}. 

For the bracketed statement [$\ldots$], we have a vector $V\psi$ with $\pi_{\mathcal{M}}(V\psi) = 1$ so the corresponding map constructed with $\psi$ has $\beta(1) = 1$. 
\end{proof}

\begin{remark}
\label{rem1}
We record here some useful observations:

\begin{enumerate}
\item[(a)] As mentioned if we were to assume that $V \left| \eta \right>$ is cyclic and separating for $\mathcal{M}$ then the last proof is much simpler, and can be found in \cite{accardi1982conditional}.  
As we have seen we can still run most of the arguments in \cite{accardi1982conditional} without the existence of such a vector. 

\item[(b)] Consider a map $\beta$ that satisfies the conditions in (iii). For any other vector $\psi \in \mathscr{H}$ we must have $\pi_{\mathcal{M}}(V\psi) \leq \beta(1)$ since $\beta(1)$
is a projector that evaluates to $1$ in the state $\omega_{V\psi}$. Thus, for the case where $\beta$ is constructed as in (ii) $\implies$ (iii) and with reference to some vector $\eta$ satisfying \eqref{stas}, then $\pi_{\mathcal{M}}(V\psi) \leq \pi_{\mathcal{M}}(V\eta)$. For example this means that if there are two vector $\eta_{1,2}$ satisfying \eqref{stas} we must have equality $\pi_{\mathcal{M}}(V\eta_1) = \pi_{\mathcal{M}}(V\eta_2)$ given (iii). In this way the map constructed in (ii) $\implies$ (iii) turns out to be unique. Although we have not shown uniqueness starting with some non-constructive $\beta$ in (iii).
If we assume that $V\eta$ is cyclic and separating for $\mathcal{M}$, then $\beta$ is always unique. 

\item[(c)] Our proof works in finite dimensions also. It is thus must be directly related to \cite{Harlow:2016vwg}, and indeed one can see similar objects floating around. In particular the use of $\Phi$ is related to the use of the reference state that is maximally entangled with the code subspace (called $\left| \phi \right>$ there).  

\end{enumerate}
\end{remark}

\bibliography{entanglement}


\end{document}